\documentclass[twocolumn,preprintnumbers,superscriptaddress,nofootinbib,aps,prd,floatfix]{revtex4}

\usepackage{amsmath,amssymb}
\usepackage{dsfont}
\usepackage{graphicx} 
\usepackage{epstopdf} 
\usepackage{slashed}
\usepackage{enumerate}

\usepackage{subfigure}
\usepackage{color}
\usepackage{multirow}
\usepackage[scientific-notation=true]{siunitx}
\usepackage{hyperref}

\usepackage{footmisc}

\usepackage{array}
\newcolumntype{L}[1]{>{\raggedright\let\newline\\\arraybackslash\hspace{0pt}}m{#1}}
\newcolumntype{C}[1]{>{\centering\let\newline\\\arraybackslash\hspace{0pt}}m{#1}}
\newcolumntype{R}[1]{>{\raggedleft\let\newline\\\arraybackslash\hspace{0pt}}m{#1}}

\newcommand{\be}{\begin{eqnarray*}}
\newcommand{\ee}{\end{eqnarray*}}
\newcommand{\gl}[1]{(\ref{#1})}

\newcommand{\bee}{\begin{eqnarray}}
\newcommand{\eee}{\end{eqnarray}}
\newcommand{\beeq}{\begin{equation}}
\newcommand{\eeeq}{\end{equation}}

\renewcommand{\vec}{\bf}

\begin{document}

\title{$hhjj$ production at the LHC}

\begin{abstract}
  The search for di-Higgs production at the LHC in order to set
  limits on Higgs trilinear coupling and constraints on new physics is one of the main
  motivations for the LHC high luminosity phase.
  Recent experimental analyses suggest that such analyses will only be
  successful if information from a range of channels is included. We
  therefore investigate di-Higgs production in association with two
  hadronic jets and give a detailed discussion of both the gluon- and
  weak boson fusion contributions, with a particular emphasis on the
  phenomenology with modified Higgs trilinear and quartic gauge
  couplings. We perform a detailed investigation of the full hadronic
  final state and find that $hhjj$ production should add sensitivity
  to a di-Higgs search combination at the HL-LHC with 3~ab$^{-1}$. Since the WBF and GF contributions are sensitive to
  different sources of physics beyond the Standard Model, we devise
  search strategies to disentangle and isolate these production
  modes. While gluon fusion remains non-negligible in WBF-type
  selections, sizeable new physics contributions to the latter can
  still be constrained. As an example of the latter point we
  investigate the sensitivity that can be obtained for a measurement
  of the quartic Higgs-gauge boson couplings.
\end{abstract}
\author{Matthew J. Dolan} \email{mdolan@slac.stanford.edu}
\affiliation{Theory Group, SLAC National Accelerator
  Laboratory,\\Menlo Park, CA 94025, USA\\[0.1cm]}
\affiliation{ARC Centre of Excellence for Particle Physics at the
  Terascale, School of Physics, University of Melbourne, 3010, Australia\\[0.1cm]}
\author{Christoph Englert} \email{christoph.englert@glasgow.ac.uk}
\affiliation{SUPA, School of Physics and Astronomy, University of
  Glasgow,\\Glasgow, G12 8QQ, United Kingdom\\[0.1cm]}
\author{Nicolas Greiner} \email{nicolas.greiner@desy.de}
\affiliation{DESY Theory Group, Notkestr. 85, D-22607 Hamburg, Germany\\[0.1cm]}
\author{Karl Nordstrom} \email{k.nordstrom.1@research.gla.ac.uk}
\affiliation{SUPA, School of Physics and Astronomy, University of
  Glasgow,\\Glasgow, G12 8QQ, United Kingdom\\[0.1cm]}
\author{Michael Spannowsky} \email{michael.spannowsky@durham.ac.uk}
\affiliation{Institute for Particle Physics Phenomenology, Department
  of Physics,\\Durham University, DH1 3LE, United Kingdom\\[0.1cm]}

\preprint{DESY 15-097, IPPP/15/38, DCPT/15/76, SLAC-PUB-16316}

\maketitle

\section{Introduction}
After the Higgs boson discovery in 2012~\cite{Chatrchyan:2012ufa} and
subsequent analyses of its properties~\cite{Hcoup}, evidence for physics
beyond the Standard Model (BSM) remains elusive. Although consistency
with SM Higgs properties is expected in many BSM
scenarios, current measurements do not fully constrain
the Higgs sector. One coupling which is currently unconstrained and has recently been subject of much interest is the Higgs
self-interaction $\sim \eta$, which is responsible for the spontaneous
breaking of electroweak gauge symmetry in the SM via the potential
\begin{equation}
  \label{eq:higgspot}
  V(H^\dagger H) =
  \mu^2 H^\dagger H + \eta (H^\dagger H)^2 \, , 
\end{equation}
with $\mu^2<0$, where $H=(0,v+h)^T/\sqrt{2}$ in unitary gauge. The
Higgs self-coupling manifests itself primarily in a destructive
interference in gluon fusion-induced di-Higgs
production~\cite{nigel,uli2,Maltoni:2014eza} through feeding into the
trilinear Higgs interaction with strength
$\lambda_{\text{SM}}=m_h\sqrt{\eta/2}=gm_h^2/(4 m_W)$ in the SM. The
latter relation can be altered in BSM scenarios, e.g. the SM coupling
pattern can be distorted by the presence of a dimension six operator
$\sim (H^\dagger H)^3$, and di-Higgs production is the only channel
with direct sensitivity to this interaction~\cite{Azatov:2015oxa}. A
 modification solely of the Higgs trilinear coupling, which is typically
invoked in di-Higgs feasibility studies, is predicted in models of
$\mu^2$-less electroweak symmetry breaking,~e.g.~\cite{womu2}.

\begin{figure*}[!t]
  \centering
  \subfigure[\label{fig:pth}]{\includegraphics[width=0.45\textwidth]{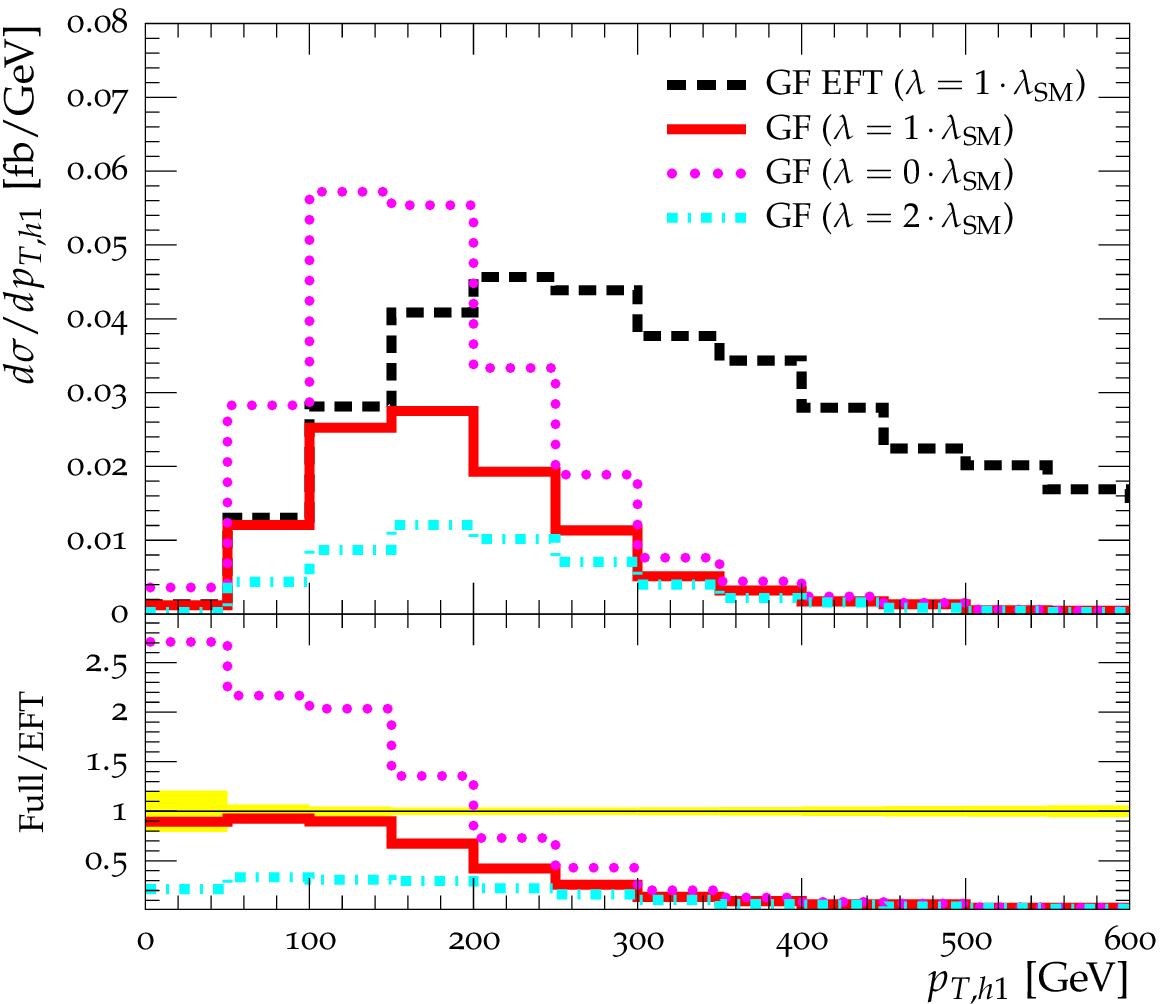}}
  \hfill
  \subfigure[\label{fig:ptj}]{\includegraphics[width=0.45\textwidth]{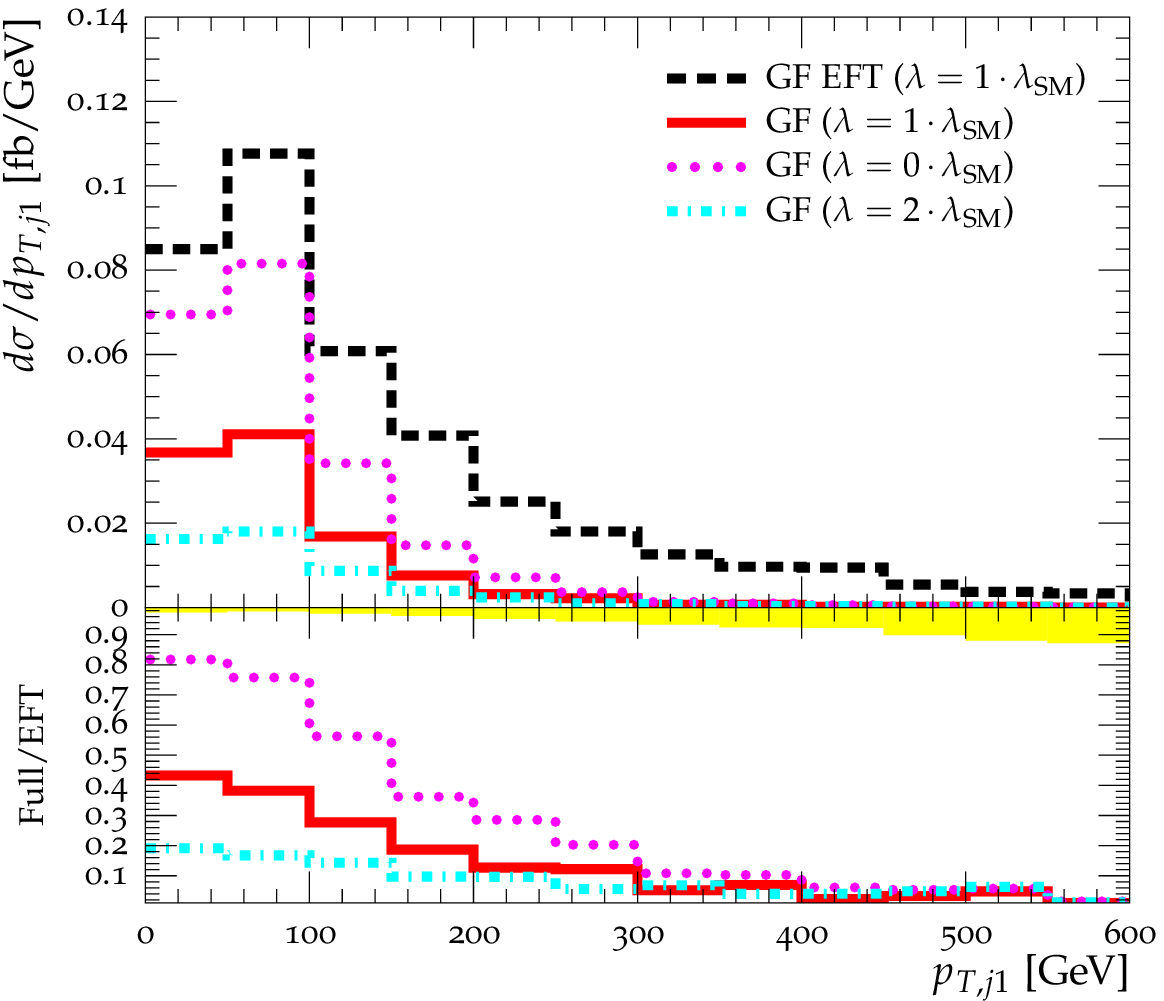}}
  \caption{\label{fig:pt} Maximum Higgs and jet transverse momenta
    for gluon fusion-induced $hhjj$ production, including the ratio of
    full theory to the effective theory calculation for three
    different values of the Higgs trilinear coupling $\lambda$.}
\end{figure*}

\begin{figure}[t!]
  \centering
  \vspace{0.3cm}
  \parbox{0.45\textwidth}{
 \includegraphics[angle=-90,width=0.22\textwidth]{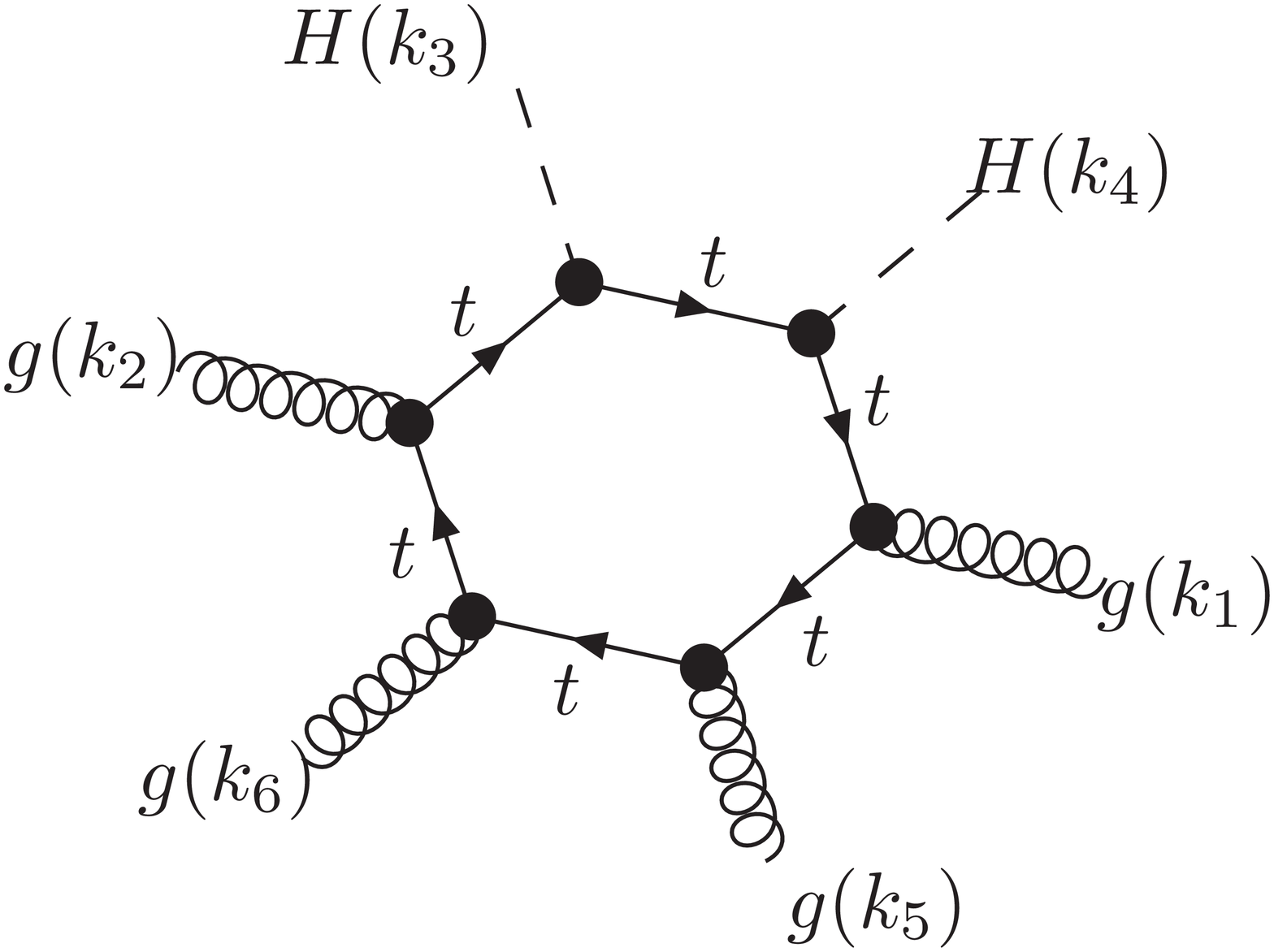}
 \includegraphics[angle=-90,width=0.22\textwidth]{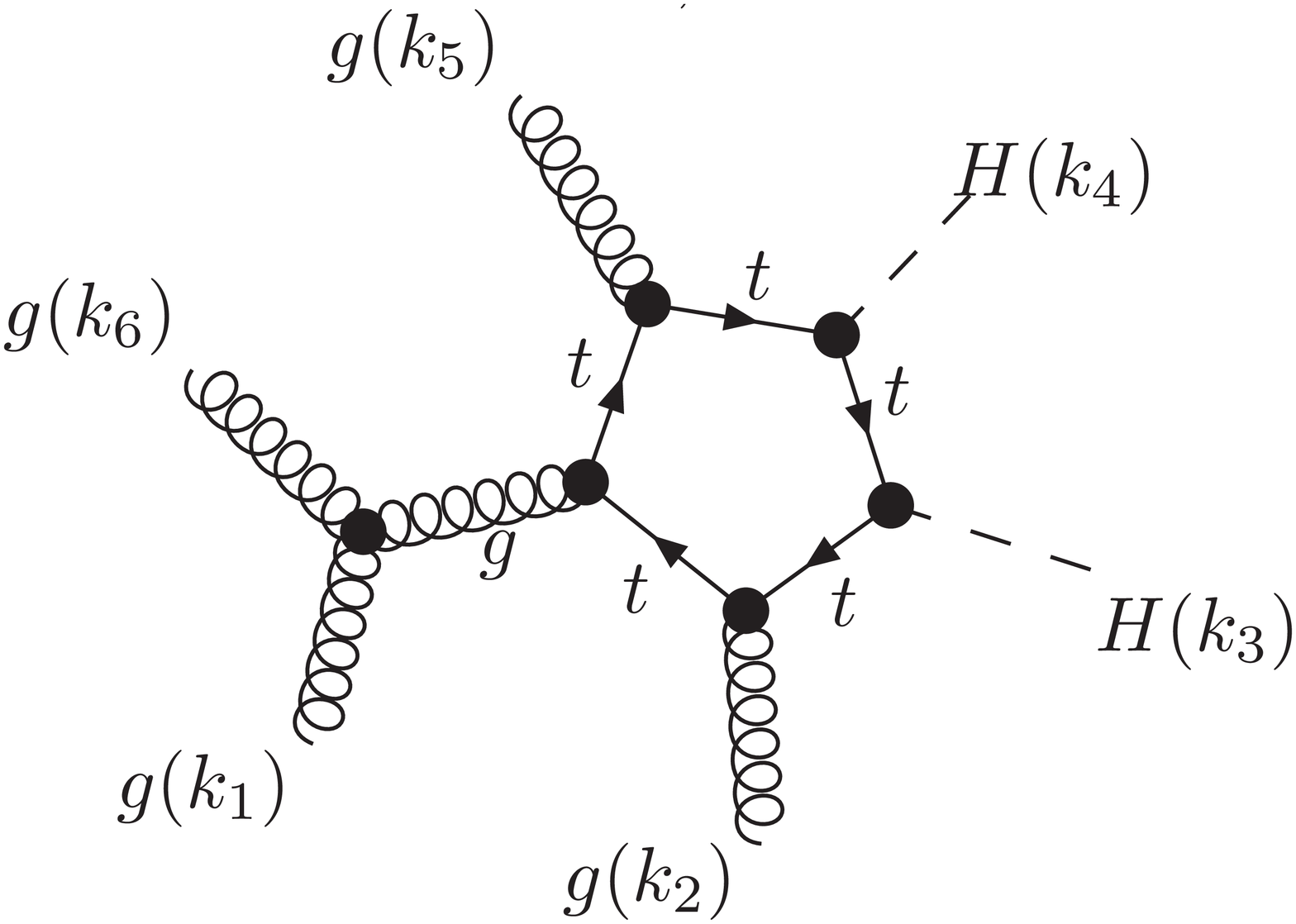}\\
 \includegraphics[angle=-90,width=0.22\textwidth]{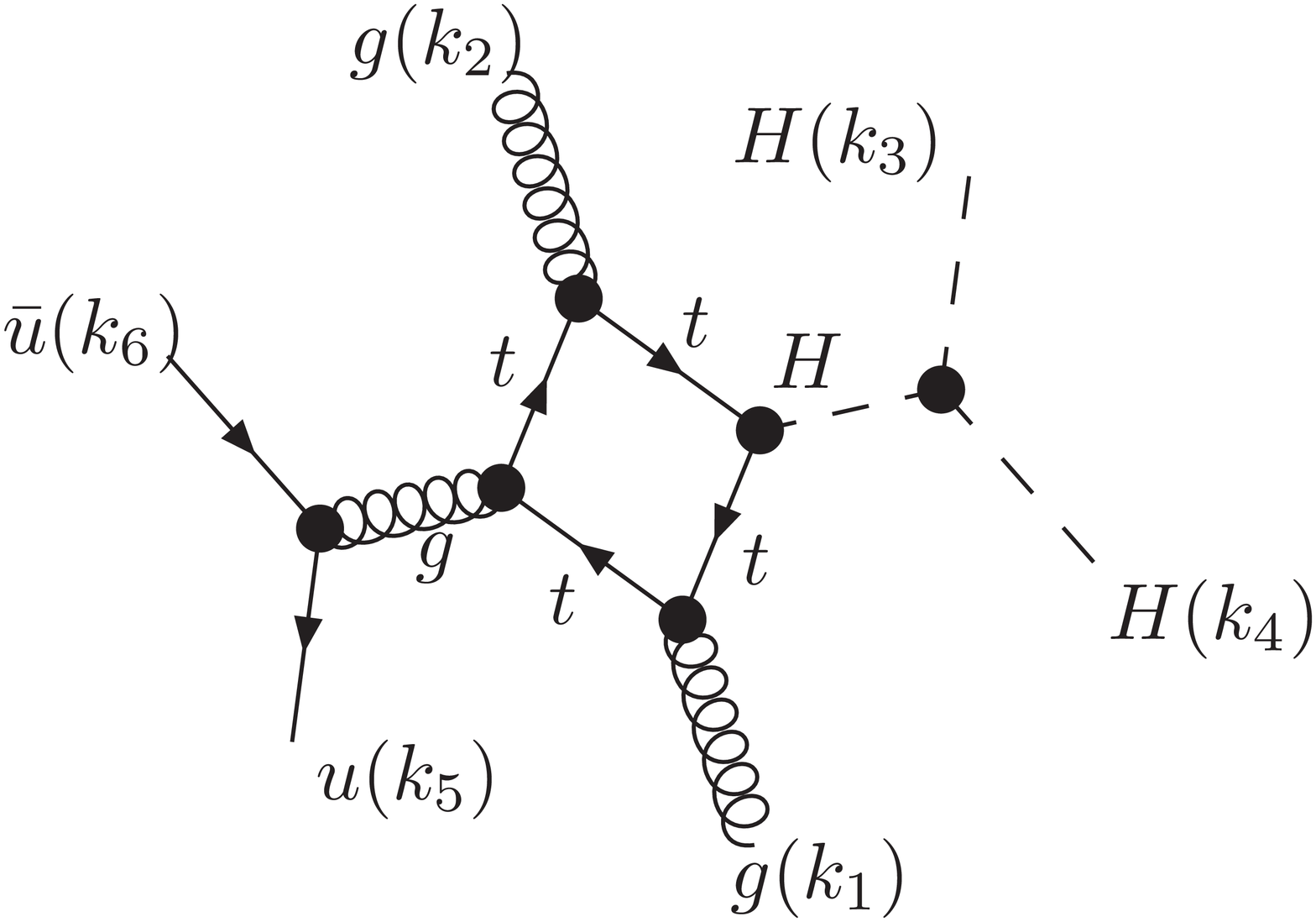}
 \includegraphics[angle=-90,width=0.22\textwidth]{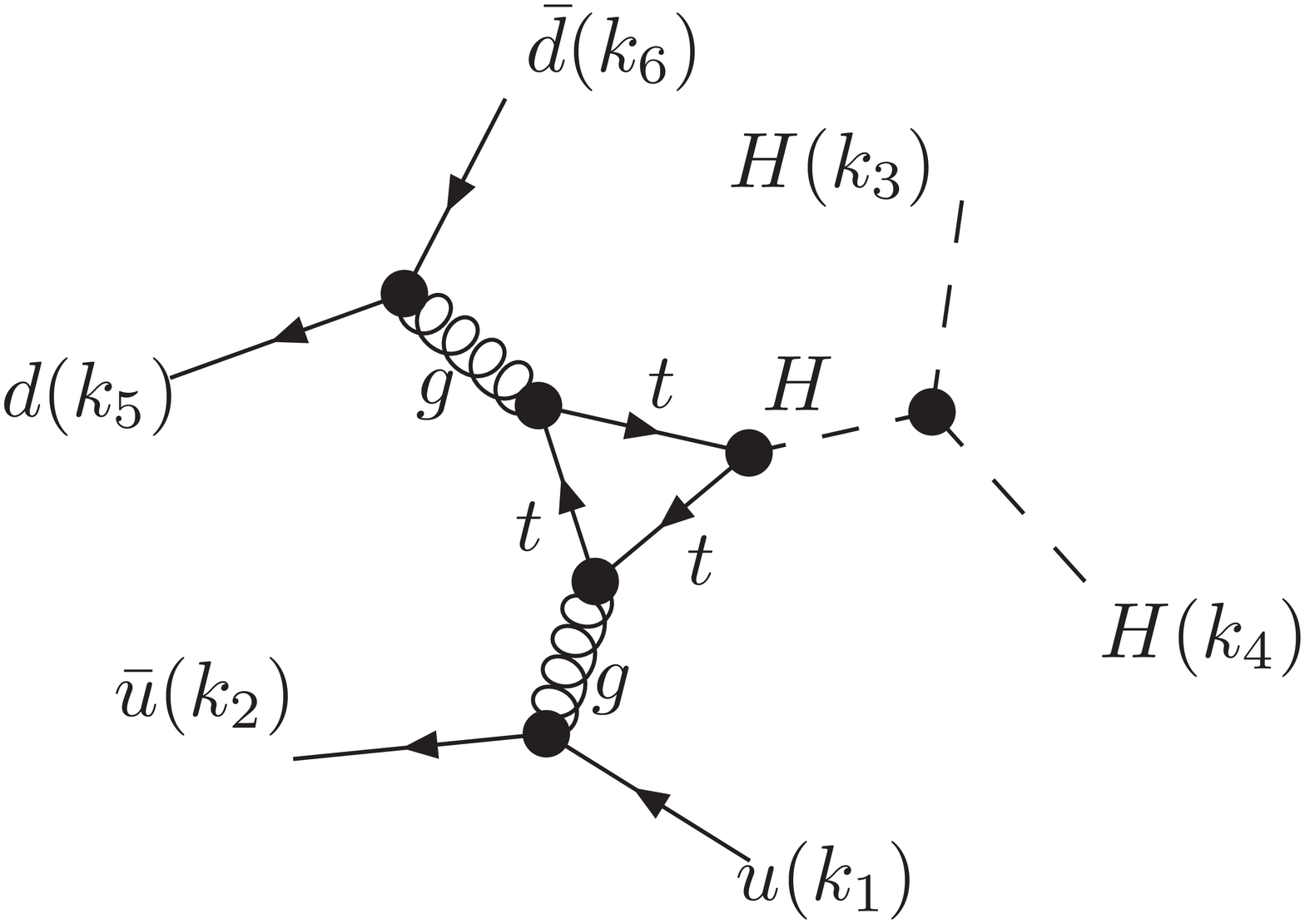}
 \caption{Sample Feynman diagrams contributing to $pp\to hhjj$ via gluon
   fusion.}
 \label{fig:sample_gf}
  \vspace{-0.5cm}
}
\end{figure}

After the Higgs discovery, analyses of the di-Higgs final state at the
high-luminosity LHC and beyond have experienced a renaissance, and
di-Higgs final states such as the $b\bar b \gamma
\gamma$~\cite{Baur:2003gp,Barger:2013jfa,Azatov:2015oxa,Barr:2014sga},
$b\bar b \tau^+\tau^-$~\cite{Baur:2003gpa,us,us2}, $b \bar b
W^+W^-$~\cite{Baur:2003gpa,Papaefstathiou:2012qe,us} and $b\bar b b
\bar b$~\cite{Baur:2003gpa,deLima:2014dta,us} channels have been
studied phenomenologically, often relying on boosted jet substructure
techniques~\cite{bdrs} (see also an
investigation~\cite{Papaefstathiou:2015iba} of rare decay channels
relevant for a 100 TeV collider). Recent analyses by ATLAS and
CMS~\cite{atlasecfa,cmsecfa} have highlighted the complexity of these
analyses and the necessity to explore different production mechanisms
to formulate constraints on the Higgs self-interactions in the
future. This program has already been initiated by feasibility
analyses of the $hhj$, $hhjj$ and $t\bar t hh$ production modes in
Refs.~\cite{us,Barr:2014sga,hhjj,tthh,Liu:2014rva}.

Di-Higgs production in association with two jets is a particularly
important channel in this regard since this final state receives
contributions from the weak boson fusion (WBF) production mode. The
phenomenological appeal of the WBF mode is twofold. Firstly, the weak
boson fusion component of $pp\to hh jj$ is sensitive to modifications
of the gauge-Higgs sector~\cite{hhjj,Brooijmans:2014eja}, which can lead to 
large cross-section enhancements. Secondly, the QCD uncertainties for
the WBF topologies are known and under theoretical
control~\cite{hhjj2,Frederix:2014hta}, such that a search for BSM
electroweak-induced deviations is not hampered by QCD
systematics. This situation is very different from QCD-induced
production~\cite{qcd}, and can be attributed to the particular
phenomenology of WBF-like processes~\cite{Figy:2003nv,hhjj1}

However, an additional source of uncertainty that was neglected until
recently~\cite{hhjj} is the correct inclusion of the gluon fusion
contribution to $pp\to hh jj$ analyses. In contrast to single Higgs
phenomenology, the correct inclusion of massive fermion
thresholds is crucial to a reliable prediction of QCD-induced $pp \to
hh jj$~\cite{us}.

Given that the cross sections in WBF $hhjj$ production are very
suppressed compared to WBF $hjj$ production (the WBF $hhjj$ cross
section is $\sim 750$ times smaller), we have to rely on the dominant
hadronic Higgs decay modes to be able to observe this final state.
This rules out one of the most crucial single Higgs WBF selection
tools - the central jet veto~\cite{dieter}. The observation of
WBF-induced $pp\to hhjj$ production is further hampered by the top
threshold in the QCD-mediated process. Since the top threshold sets
the scale of the di-Higgs subsystem, an analysis that tries to retain
as many low $p_T$ Higgs bosons as possible leads to a QCD contribution
that dominates over the WBF component when minimal WBF-like cut
requirements are imposed~\cite{hhjj}.

In this paper we extend the discussion of Ref.~\cite{hhjj} in a number
of directions. We first perform a detailed comparison of
EFT-approaches to QCD-mediated $pp\to hhjj$ against a calculation
keeping the full mass dependencies of top and bottom quarks in
Sec.~\ref{sec:gf}. We compare the QCD-induced $pp\to hhjj$
phenomenology to the WBF signature in Sec.~\ref{sec:wbf} before we
discuss general approaches to isolate the signal from the dominant top
backgrounds in a hadron level analysis in Sec.~\ref{sec:tt}. This sets
the stage for a discussion about the prospects to isolate the WBF and
GF components in Secs.~\ref{sec:isolgf} and \ref{sec:isolwbf},
followed by a study on constraining $VV hh$ couplings using the WBF
induced signal in Section~\ref{sec:limitzeta}. We focus on collisions
with $14$ TeV throughout.

\section{The gluon fusion contribution}
\label{sec:gf}
\subsection{Finite top mass effects}

It is well known that effective field theory approximations in the
$m_t\to \infty$ limit cannot be invoked to study di-Higgs final states
at colliders in a reliable way due to the effects of top-quark
threshold~\cite{uli2,Dawson:2015oha}. Further, the breakdown of the $m_t\to
\infty$ approximation is worsened in the presence of additional jet
emission~\cite{us,Maierhofer:2013sha}. Finite $m_t$ effects must
therefore be considered for all QCD di-Higgs production channels,
which will be required to set the best limits on the Higgs
self-coupling or formulate a realistic estimate of the GF contribution
in a WBF-like selection.

\begin{table}[!t]
 \begin{tabular}{lccc}
  \toprule
  $\lambda$ \hspace{1cm}  & \hspace{0.25cm}	$0 \cdot \lambda_\text{SM}$ [fb] \hspace{0.25cm}  & \hspace{0.25cm} $1 \cdot \lambda_\text{SM}$ [fb]\hspace{0.25cm}  & \hspace{0.25cm} $2 \cdot \lambda_\text{SM}$ [fb] \\
  \botrule
   GF	&	10.73					     & 5.502					    & 2.669		 \\
   WBF  &	4.141					     & 2.010					    & 0.9648		\\
   \botrule
 \end{tabular}
 \caption{Cross section normalisations for the GF and WBF
   samples at 14~TeV, for details see text. The WBF normalisation follows
   from~\cite{hhjj2} and includes higher order QCD effects.}
 \label{tab:xsnormalisation}
\end{table}

The computational challenges in QCD-mediated $hhjj$ production are
significant, with the gluon-fusion channels particularly time
consuming even when using state-of-the-art techniques. The standard
method of simulating a differential cross section from unweighted
events is not feasible in this case, and we instead use a reweighting
technique that is exploited in higher order calculations and
experimental analyses~(see e.g. \cite{Aad:2013nca}).

We generate GF $hhjj$ events by implementing the
relevant higher dimensional operators in the $m_t\to \infty$ limit
obtained by expanding the low-energy effective theory~\cite{kniehl}
\begin{equation}
  \mathcal{L}_\text{eff} = -\frac{1}{4} \frac{\alpha_S}{3 \pi} 
  G^a_{\mu \nu} G^{a\, \mu \nu} \log(1 + h/v)
  \label{eq:efft}
\end{equation}
in \textsc{MadEvent} v5.1~\cite{Alwall:2011uj} using the
\textsc{FeynRules/Ufo}~\cite{Degrande:2011ua} framework.\footnote{The
  effective theory implementation can be modified in the sense that
  only one effective vertex insertion is allowed. This is gives only a
  mild $\sim 10\%$ effect in the tail of the distribution, and is not
  relevant for an order one EFT/full theory rescaling, see below.}
This allows us to sample a weighted set of events that we subsequently
feed into our analysis solely depending on their final state
kinematics. If an event passes the selection requirements of a certain
search region, we correct for the full mass dependence using a
reweighting library based on \textsc{GoSam} package~\cite{gosam} at
this stage. The reweighting employs exactly the same matrix elements
used for the event generation and the trilinear coupling is steered
through a modification of the {\sc{GoSam}} matrix element,
i.e. variations of the trilinear coupling are part of the
reweighting. A selection of Feynman diagrams which contribute to the
gluon fusion signal are illustrated in Fig.~\ref{fig:sample_gf}. The
{\sc{GoSam}} code used for the reweighting is based on a Feynman
diagrammatic approach using {\sc{QGRAF}}~\cite{qgraf} and
{\sc{FORM}}~\cite{form} for the diagram generation, and
{\sc{Spinney}}~\cite{spinney}, {\sc{Haggies}}~\cite{haggies} and
{\sc{FORM}} to write an optimised fortran output. The reduction of the
one-loop amplitudes was done using {\sc{Samurai}}~\cite{samurai},
which uses a $d$-dimensional integrand level decomposition based on
unitarity methods~\cite{unitarity}. The remaining scalar integrals
have been evaluated using {\sc{OneLoop}}~\cite{oneloop}. Alternative
reduction techniques can be used employing {\sc{Ninja}}~\cite{ninja}
or {\sc{Golem95}}~\cite{golem95}. To validate the reweighting
procedure we regenerated the code that has been used in~\cite{hhjj} 
with the improvements that became available within {\sc{GoSam 2.0}},
in particular improvements in code optimisation and in the reduction
of the amplitudes. For the reduction we used
{\sc{Ninja}}, which employs an improved reduction algorithm based
on an Laurent expansion of the integrand. This leads to substantial
improvements in both speed and numerical stability compared to the 
previous version. We combined the
code with a phase space integration provided by
\textsc{MadEvent}~\cite{Maltoni:2002qb}.  Further substantial speed-up
has been obtained by Monte Carlo sampling over the helicities rather
then performing the helicity sum. This enabled us to perform a full
phase space integration and we found full agreement within the 
statistical uncertainties between the result obtained from reweighting
and the result from the full phase space integration.

\begin{figure*}[t!]
  \subfigure[\label{fig:mjj}]{\includegraphics[width=0.45\textwidth]{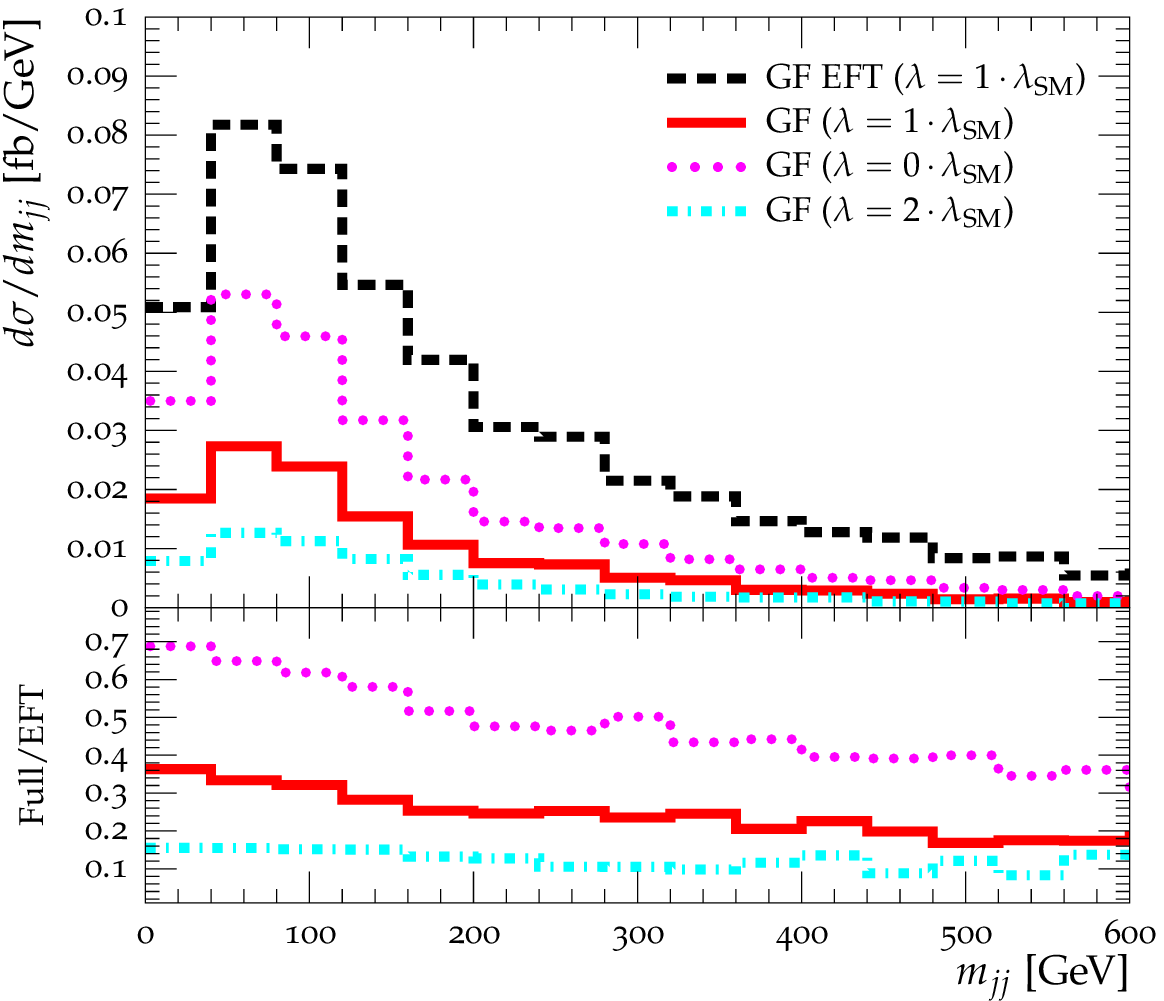}}
  \hfill
  \subfigure[\label{fig:etajj}]{\includegraphics[width=0.44\textwidth]{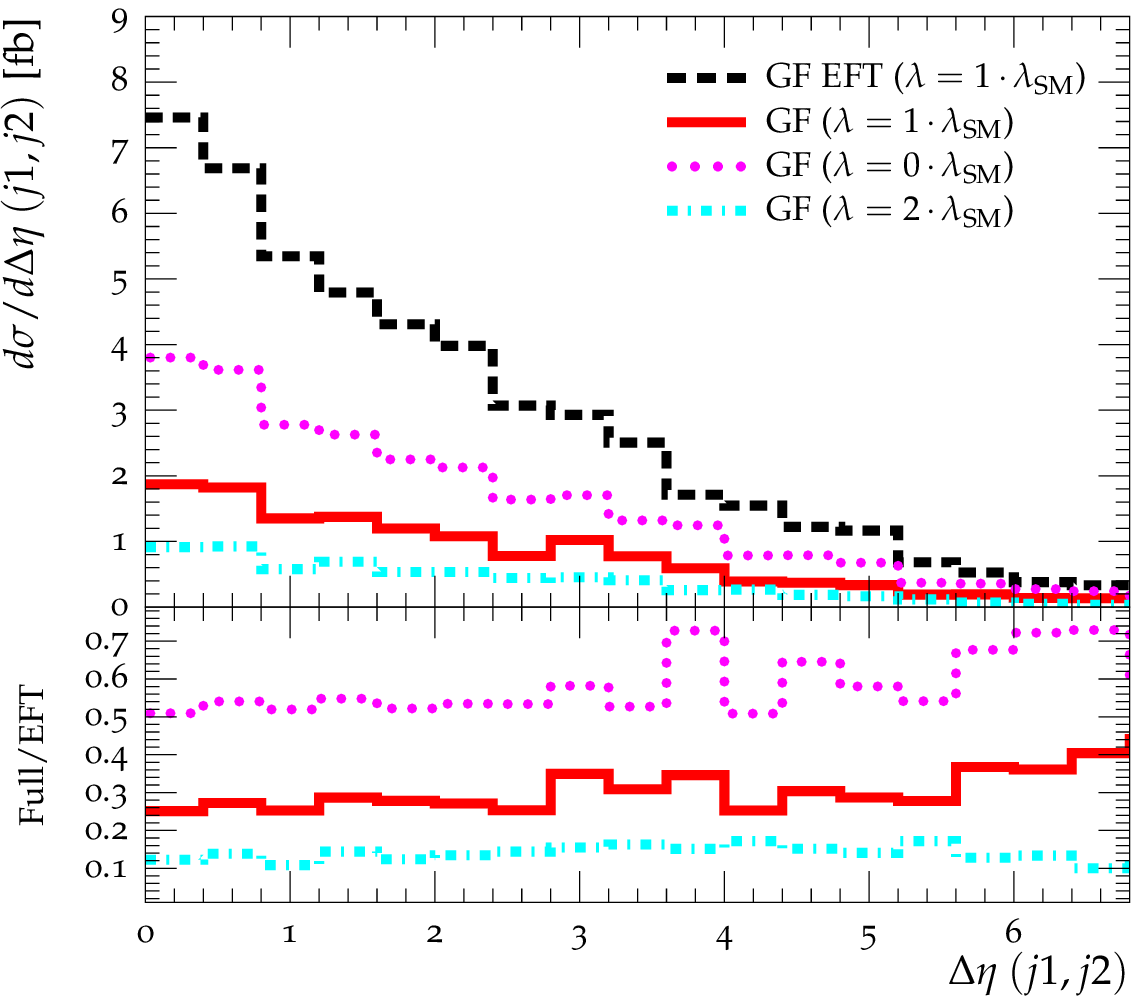}}
  \caption{\label{fig:jets} Invariant mass and pseudo-rapidity
    distributions of the jet system in QCD-mediated $hhjj$
    production. We show the  effective theory and full theory results for
    three values of the trilinear Higgs coupling, applying only generator-level cuts of $p_{T,j} \ge 20$ GeV
    and $|\eta_j| < 5$.}
\end{figure*}

\begin{figure*}[t!]
  \subfigure[\label{fig:mhh}]{\includegraphics[width=0.45\textwidth]{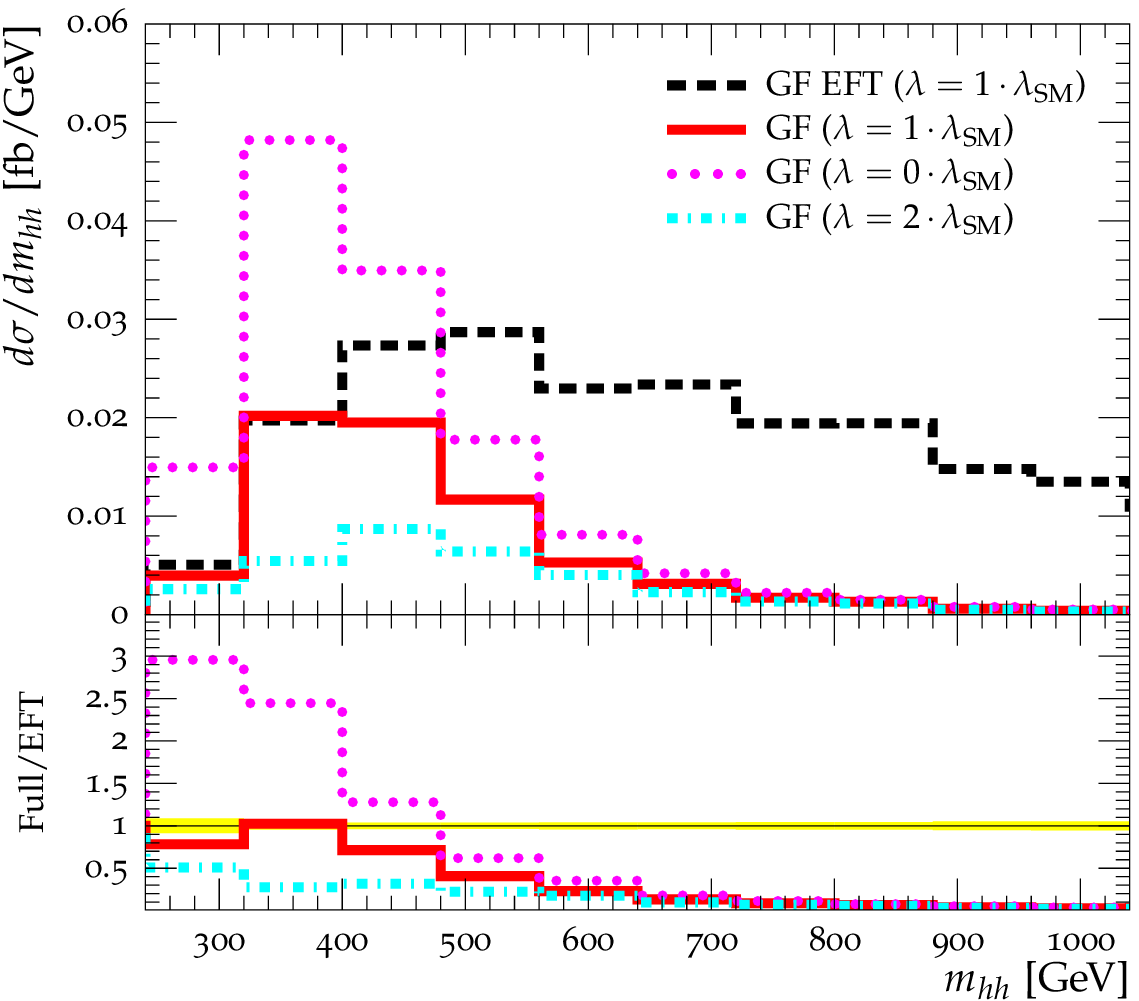}}
  \hfill
  \subfigure[\label{fig:etahj}]{\includegraphics[width=0.45\textwidth]{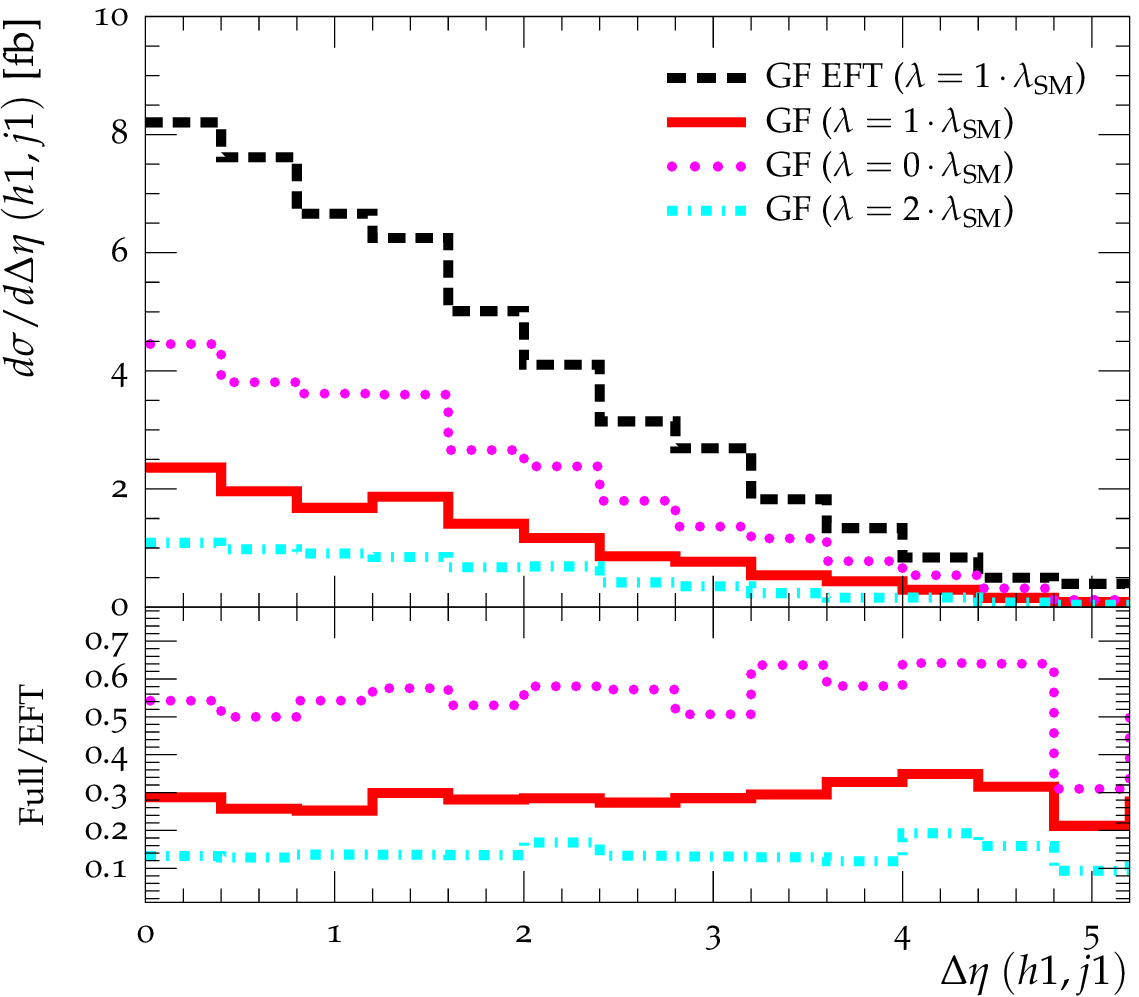}}
  \caption{\label{fig:hjet} Invariant mass and pseudo-rapidity distributions of the leading jet and di-Higgs system in QCD-mediated $hhjj$ production. We show the  effective theory and full theory results for
    three values of the trilinear Higgs coupling, applying only generator-level cuts of $p_{T,j} \ge 20$ GeV
    and $|\eta_j| < 5$.}
\end{figure*}

\subsection{Phenomenology of QCD-mediated $hhjj$ production}

Top thresholds are particularly prominent in the di-Higgs invariant
mass distribution, which is thus well suited
to benchmark the relation of the finite $m_t$ limit to the effective theory of Eq.~\gl{eq:efft}. Other
observables constructed from the six particle final state are also
relevant when performing a targeted phenomenological analysis, and we
discuss both these and the  phase space dependent parton-level reweighting in detail
in the following.

In Figures~\ref{fig:pt}, \ref{fig:jets}, and \ref{fig:hjet} we show a
selection of $hhjj$ final state observables for inclusive cuts
$p_{T,j}>20~\text{GeV}$ and $|\eta_j|< 5$, no cuts on Higgs bosons are
imposed. We label Higgs bosons and jets according to their hardness,
i.e. $p_{T,h1}>p_{T,h2}$ and $p_{T,j1}>p_{T,j2}$. The cross sections
are given in Tab.~\ref{tab:xsnormalisation}. The inclusive gluon
fusion cross section is about 2.5 times larger than the WBF cross
section approximately independent of the value of the Higgs trilinear
coupling.

As previously established in~\cite{uli2,us,hhjj} the di-Higgs system
is badly modelled by the effective theory which under- and overshoots
the full theory cross section at low and high momenta
respectively. For $pp\to hhjj$ this is a qualitatively similar
behaviour compared the $pp\to hh(j)$ production: The $m_{hh}$
distribution is the crucial observable which parametrises the finite
top quark mass effects. The EFT describes the low maximum transverse
Higgs momenta $p_{T,h1}$ reasonably well, as shown in
Fig.~\ref{fig:pth}. The jet emission on the other hand integrates over
a considerable range of $m_{hh}$, and the ratio of full theory vs
effective theory smaller than one for the finite $m_t$ limit produces
a smaller integrated cross section than the $m_t\to \infty$ limit for
the jet kinematics.

Considering just the dijet system in Fig.~\ref{fig:jets}, we observe
that the jet kinematics is not severely impacted by the reweighting
procedure upon marginalising over the di-Higgs kinematics. The phase
space dependence of the dijet invariant mass Fig.~\ref{fig:mjj} is
relatively mild aside from the total rescaling of the inclusive cross
sections, and the ratio for the pseudo-rapidity distribution of
the jets is nearly flat, Fig.~\ref{fig:etajj}. This is also true for
the azimuthal angle difference $\Delta\phi_{jj}$. The angular
distributions of the leading members of the jet-Higgs system are
relatively mildly impacted by the reweighting too
Fig.~\ref{fig:etahj}. This agrees with the $m_{hh}$ being the
observable most sensitive to the top threshold (as in $pp\to hh(j)$),
and is also supported by the larger impact of the reweighting of
$m_{hh}$ in Fig.~\ref{fig:mhh}. A reweighting based on $m_{hh}$ to
  correct for finite top mass effects suggests itself for future
  analyses as a time-saving approach with reasonable accuracy.

\section{The weak boson fusion contribution}
\label{sec:wbf}

\begin{figure}[!t]
   \vspace{1.2cm}
  \parbox{0.45\textwidth}{
    \includegraphics[width=0.22\textwidth]{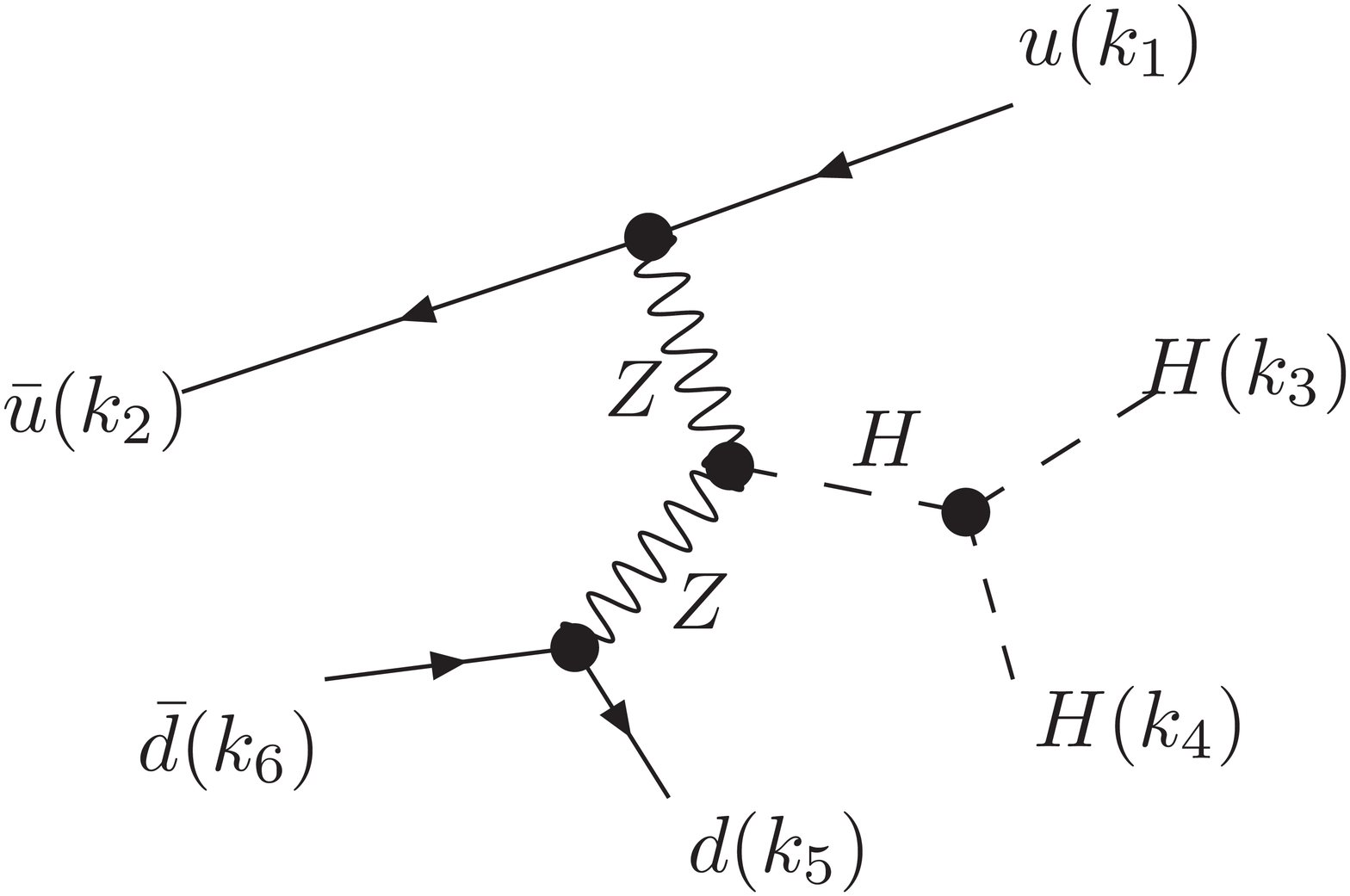}
    \includegraphics[width=0.22\textwidth]{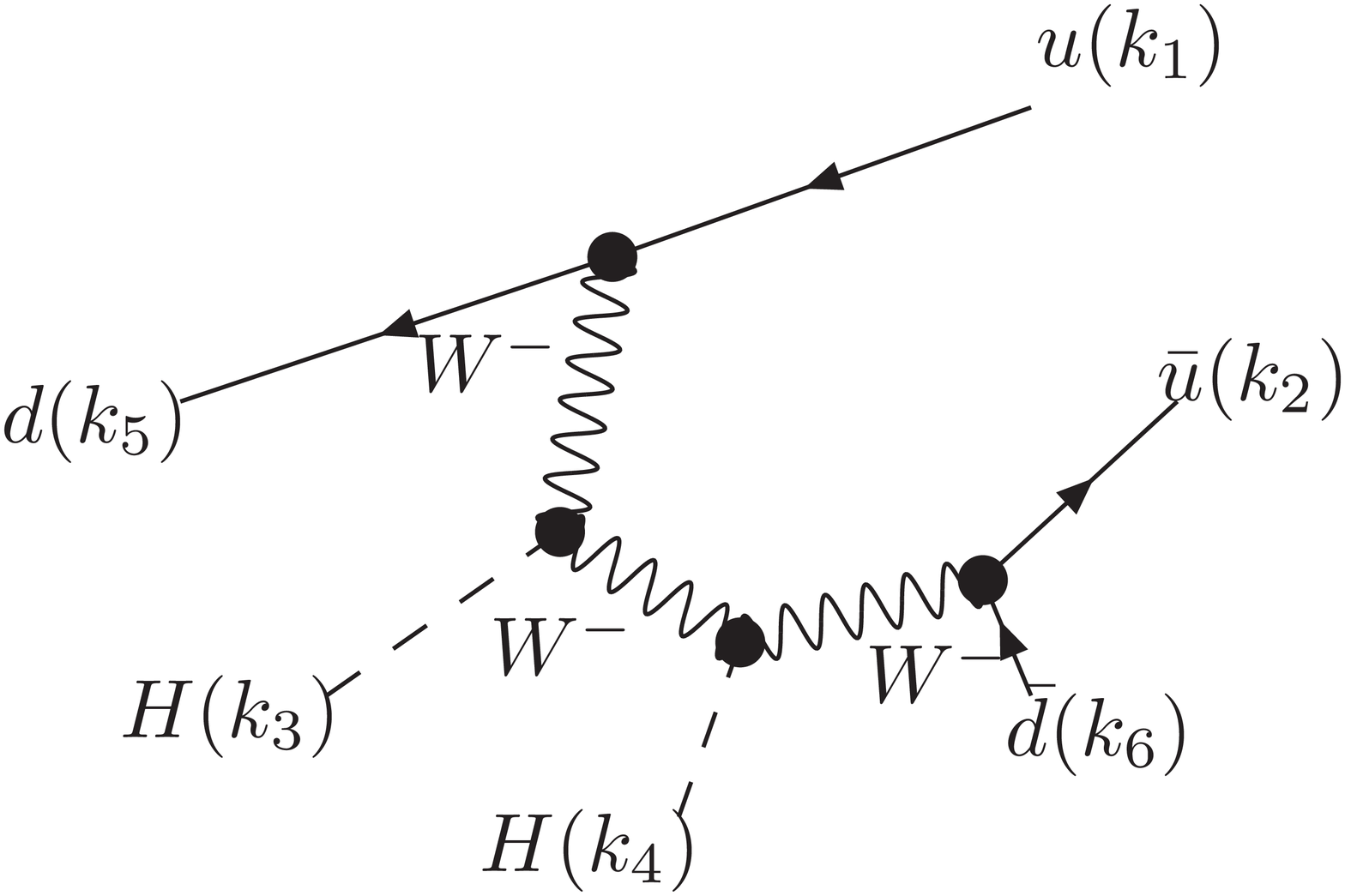}\\
    \includegraphics[width=0.22\textwidth]{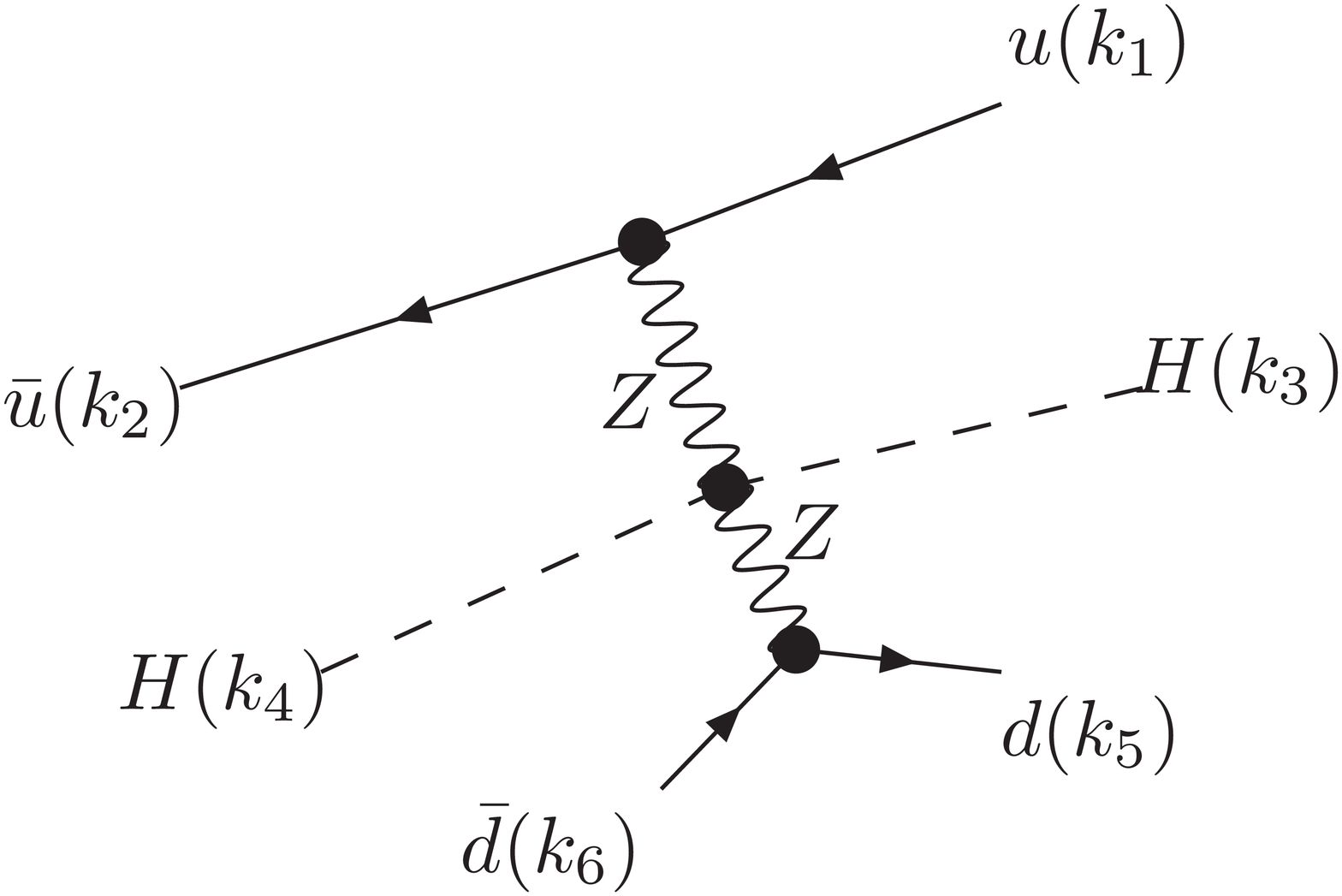}
    \includegraphics[width=0.22\textwidth]{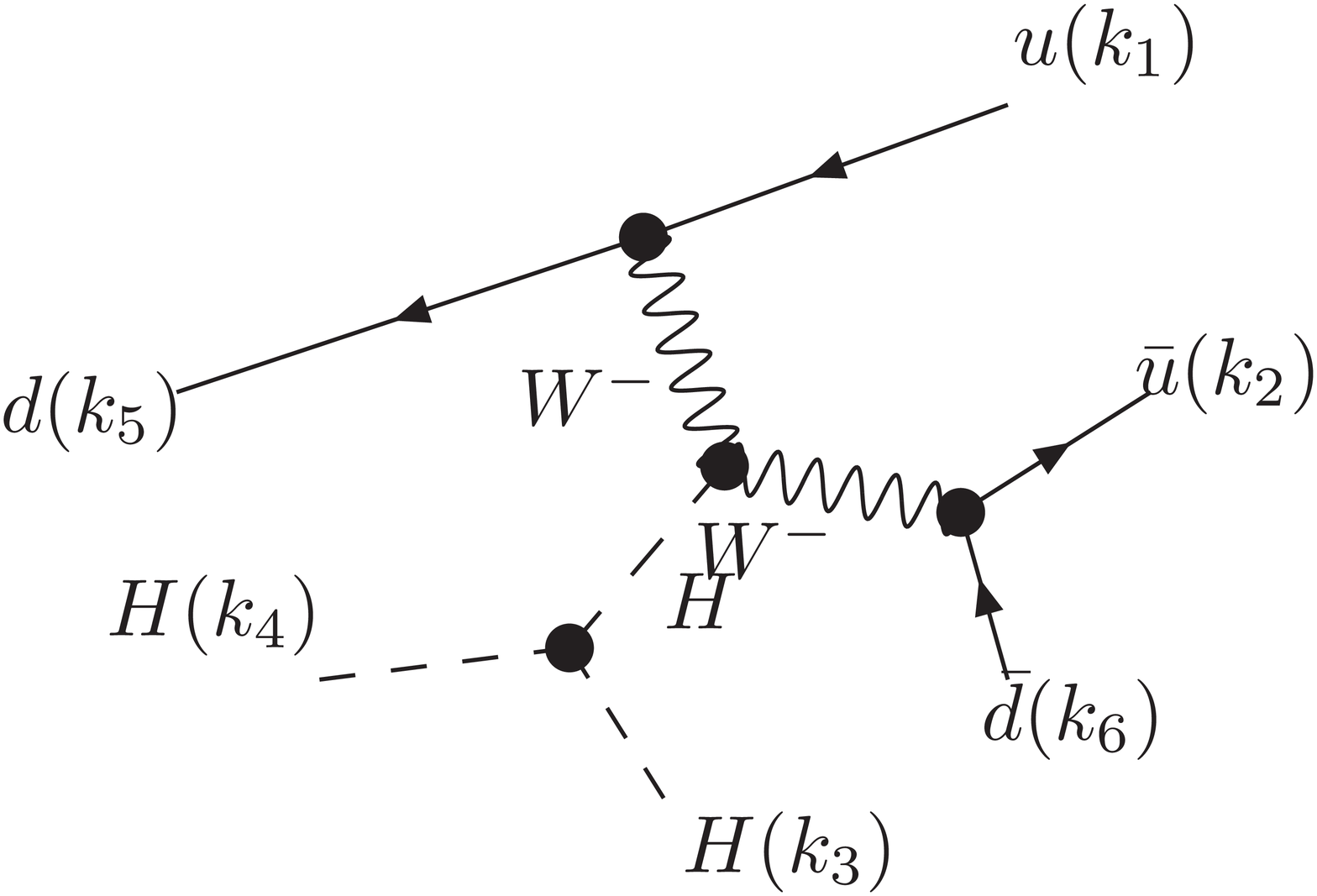}
  }
  \caption{Sample Feynman diagrams contributing to $pp \to hhjj$ in weak boson
    fusion. \label{fig:sample_wbf}}
\end{figure}

The weak boson fusion contribution to $pp\to hhjj$ has received
considerable attention recently and precise higher-order QCD
corrections have been provided
in~\cite{hhjj1,hhjj2,Frederix:2014hta}. Due to the sensitivity of the
WBF contribution to both the trilinear coupling and the quartic $VVhh$
($V=W,Z,\gamma$), as shown in the Feynman diagrams in
Fig.~\ref{fig:sample_wbf}, weak boson fusion to two Higgs bosons can,
in principle, provide complementary information about BSM physics
which remains uncaptured in $pp\to hh(j)$ and $pp\to t\bar t
hh$~\cite{Brooijmans:2014eja}.

\begin{figure}[t!]
\centering
\subfigure[\label{fig:wbfeta}]{\includegraphics[width=0.45\textwidth]{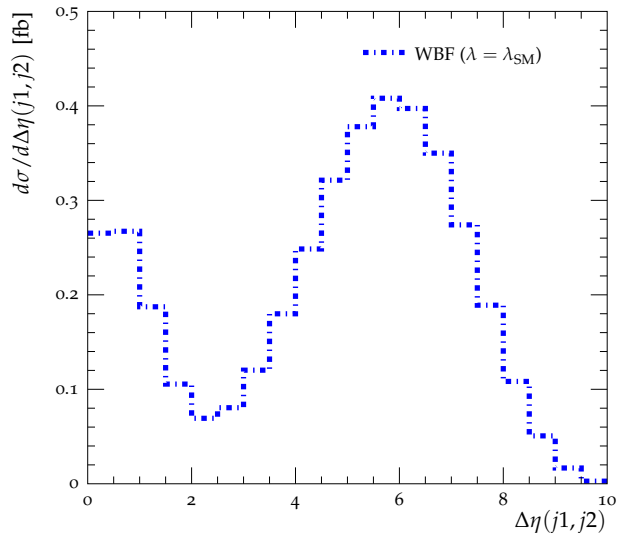}} \\[0.1cm]
\subfigure[\label{fig:gfwbfmhh}]{\includegraphics[width=0.45\textwidth]{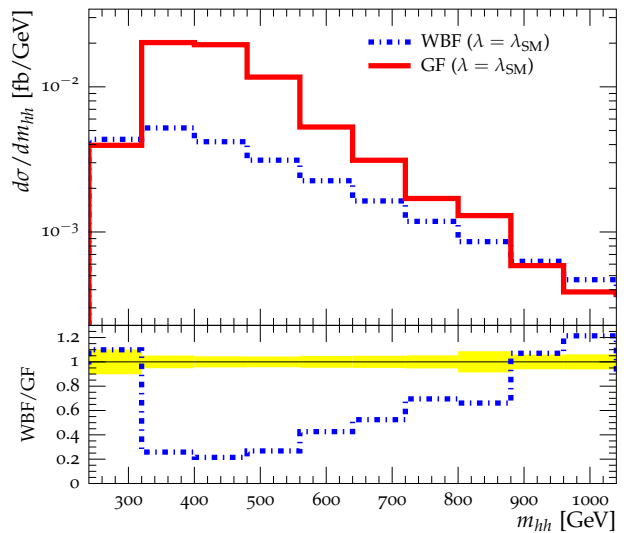}}
\caption{The $\Delta \eta(j1,j2)$ distribution of the weak boson fusion
  contribution at parton-level (a) and the $m_{hh}$ distribution of
  the weak boson fusion and gluon fusion contributions compared with
  correct cross section normalisation (b), both satisfying generator-level
  cuts of $p_{T,j} \ge 20$ GeV and $|\eta_j| < 5$.}
\label{fig:detapartonlvl}
\end{figure}

We generate WBF samples with varying $\lambda$ using \textsc{MadEvent}
v4~\cite{Alwall:2007st} and normalise the cross section to NLO
accuracy~\cite{hhjj2}.  The WBF $hhjj$ contribution shares the QCD properties
of WBF $hjj$ production \cite{Figy:2003nv} which means it shares
the distinctive $\Delta \eta(j1,j2)$ distribution shown in
Fig.~\ref{fig:wbfeta}: To produce the heavy di-Higgs pair we probe the
initial state partons at large momentum fractions. This together with
a colour-neutral $t$-channel exchange of the electroweak
bosons~\cite{gaporig} (see also \cite{gaporig2}) leads to energetic
back-to-back jet configurations at large rapidity separation and
moderate transverse momenta with a centrally produced Higgs pair. The
production of an additional Higgs boson in comparison to single Higgs
production via WBF leads to a cross section reduction by three orders
of magnitude (see Tab.~\ref{tab:xsnormalisation}) in the SM. Such a
small inclusive production cross section highlights the necessity of
considering dominant Higgs decay channels such as $h\to b \bar b$ and
$h\to \tau^+\tau^-$ and the non-availability of central jet vetos
\cite{dieter} as a means to control the background {\it{and}} GF
contribution in a targeted analysis as a consequence.

The gluon fusion contribution is bigger by a factor of 2.5 than the
WBF component of $hhjj$ production, however, with increasing invariant
di-Higgs mass the WBF contribution is enhanced relative to GF
production as a consequence of the suppression above the $2m_t$
threshold, as shown in Fig.~\ref{fig:gfwbfmhh}.

\begin{table*}[!t]
\renewcommand\arraystretch{1.1}
    \begin{tabular}{llllr}
    \toprule
    Cut setup      					&  Base Selection   [fb]\hspace{0.1cm}	& GF Selection [fb]\hspace{0.1cm}& WBF Selection [fb] \hspace{0.1cm} &   Normalisation$*$ [fb] \\
    \botrule
    GF	($\lambda = 1 \cdot \lambda_\text{SM}$)	& \hspace{0.1cm} \num{0.01396}		& \hspace{0.1cm}  \num{0.005722} 		& \hspace{0.1cm}  \num{0.0005378}      &	\num{0.4013}	\\
    GF	($\lambda = 0 \cdot \lambda_\text{SM}$)	& \hspace{0.1cm} \num{0.02562}       	& \hspace{0.1cm}  \num{0.008122}    		& \hspace{0.1cm}  \num{0.0008767}      &	\num{0.7831}	\\
    GF	($\lambda = 2 \cdot \lambda_\text{SM}$)	& \hspace{0.1cm} \num{0.007167}      	& \hspace{0.1cm}  \num{0.003906}    		& \hspace{0.1cm}  \num{0.0003034}     &	\num{0.1947}	\\
    WBF	($\lambda = 1 \cdot \lambda_\text{SM}$)	& \hspace{0.1cm} \num{0.003292}      	& \hspace{0.1cm}  \num{0.0004999}   		& \hspace{0.1cm}  \num{0.001485}      & 	\num{0.1466}	\\
    WBF	($\lambda = 0 \cdot \lambda_\text{SM}$)	& \hspace{0.1cm} \num{0.007706}      	& \hspace{0.1cm}  \num{0.0007154}   		& \hspace{0.1cm}  \num{0.002820}      &	\num{0.3020}	\\
    WBF	($\lambda = 2 \cdot \lambda_\text{SM}$)	& \hspace{0.1cm} \num{0.001103}      	& \hspace{0.1cm}  \num{0.0001815}   		& \hspace{0.1cm}  \num{0.0003912}     &	\num{0.07037}   \\
    $t\bar{t}jj$					& \hspace{0.1cm} \num{5.712}	      	& \hspace{0.1cm}  \num{0.03390}   		& \hspace{0.1cm}  \num{0.01801}       &	\num{10130}	\\
    $t\bar{t}h$						& \hspace{0.1cm} \num{0.06229}       	& \hspace{0.1cm}  \num{0.007047}   		& \hspace{0.1cm}  \num{0.00005658}     & 	\num{38.62}	\\
    $Zhjj$						& \hspace{0.1cm} \num{0.005118}      	& \hspace{0.1cm}  \num{0.001278}   		& \hspace{0.1cm}  \num{0.0001026}     & 	\num{47.37}	\\
    $ZZjj$						& \hspace{0.1cm} \num{0.001171}  	& \hspace{0.1cm}  \num{0.00006659}   		& \hspace{0.1cm}  \num{0.0000007639}	&	\num{225.7}	\\
    $ZWWjj$						& \hspace{0.1cm} \num{0.00001888}       & \hspace{0.1cm}  \num{0.000005461}   		& \hspace{0.1cm}  \num{0.0000002039}	&	\num{0.5368}  \\
    total background					& \hspace{0.1cm} \num{5.781}       	& \hspace{0.1cm}  \num{0.04230}   		& \hspace{0.1cm}  \num{0.01817}	&	-	\\
    \botrule    
    $S/B$	($\lambda = 1 \cdot \lambda_\text{SM}$) 			& \hspace{0.3cm} 1/335.1       	& \hspace{0.3cm} 1/6.799	& \hspace{0.3cm}  1/8.983          \\
    $S/B$	GF$^\dagger$	 ($\lambda = 1 \cdot \lambda_\text{SM}$)	& \hspace{0.3cm} 1/414.3       	& \hspace{0.3cm} 1/7.480 	& \hspace{0.3cm}  1/36.55         \\
    $S/B$	WBF$^\dagger$	($\lambda = 1 \cdot \lambda_\text{SM}$)		& \hspace{0.3cm} 1/1760       	& \hspace{0.3cm} 1/96.06 	& \hspace{0.3cm}  1/12.60          \\
 \botrule
    $S/\sqrt{\text{B}}$ (3 ab$^{-1}$, $\lambda = 1 \cdot \lambda_\text{SM}$)		& \hspace{0.3cm} 0.3930       & \hspace{0.3cm} 1.657	& \hspace{0.3cm}  0.8219        \\
 \botrule
 $*$ branchings included in normalisation \\
 $^\dagger$ considering only this as signal
     \end{tabular}
     \caption{Cross sections for the two sources of signal, and backgrounds, after the various selections described in the text are applied, together with various measures of significance in the bottom four rows.}
 \label{tab:xstable}
\end{table*}

Since we cannot rely on vetoing hadronic activity in the central part
of the detector, a potential discrimination of GF from WBF needs to be
built on the following strategy, which we will investigate in
Sec.~\ref{sec:tt}:

\begin{itemize}
  \item To isolate the di-Higgs (WBF+GF) signal we can exploit the relative
    hardness of the di-Higgs pair which peaks around $\sim 2m_t$. Such
    hard events are less likely to be produced by (ir)reducible
    backgrounds.
  \item Focussing on large $m_{hh}$ we can enhance WBF over GF by
    stringent cuts on the jet rapidity separation. This will also
    imply a significant decrease of QCD-dominated backgrounds.
  \item By explicitly allowing central jet activity, we can exploit
    the colour correlation differences in WBF vs GF to further purify
    our selection. Since colour flow is tantamount to energy flow in the
    detector,  event shapes are particularly well-suited observables
    for unravelling the colour correlations in the final state once
    the reconstructed di-Higgs pair has been removed\footnote{A
      detailed discussion of event shapes at hadron colliders can be
      found in~\cite{Banfi:2010xy}.}. This strategy was first proposed
    for single Higgs
    analyses in \cite{Englert:2012ct} (see also
    \cite{plehn}).
\end{itemize}

\section{Taming the background}
\label{sec:tt}
For our hadron-level analysis we shower our signal samples with
\textsc{Herwig++}~\cite{Bahr:2008pv} and generate backgrounds as
follows: $t\bar{t}jj$, $t\bar{t}h$, $Zhjj$, and $ZZjj$ with
\textsc{Sherpa}~\cite{Gleisberg:2008ta}, and $ZWWjj$ with
\textsc{MadEvent} v5. We find the dominant backgrounds to be
$t\bar t jj$ and $t\bar t h$ production, for which next-to-leading
order results are available~\cite{ttjjnlo,tthnlo}
and we use inclusive $K$ factors $K_{t\bar t jj}\simeq 1$ and $K_{t\bar t
  h}\simeq 1.5$ to estimate the higher order contributions
to these backgrounds. Higgs branching ratios are set to the values
agreed upon by the Higgs Cross Section Working
Group~\cite{Heinemeyer:2013tqa}.

\begin{figure*}[t!]
\centering
\includegraphics[width=0.45\textwidth]{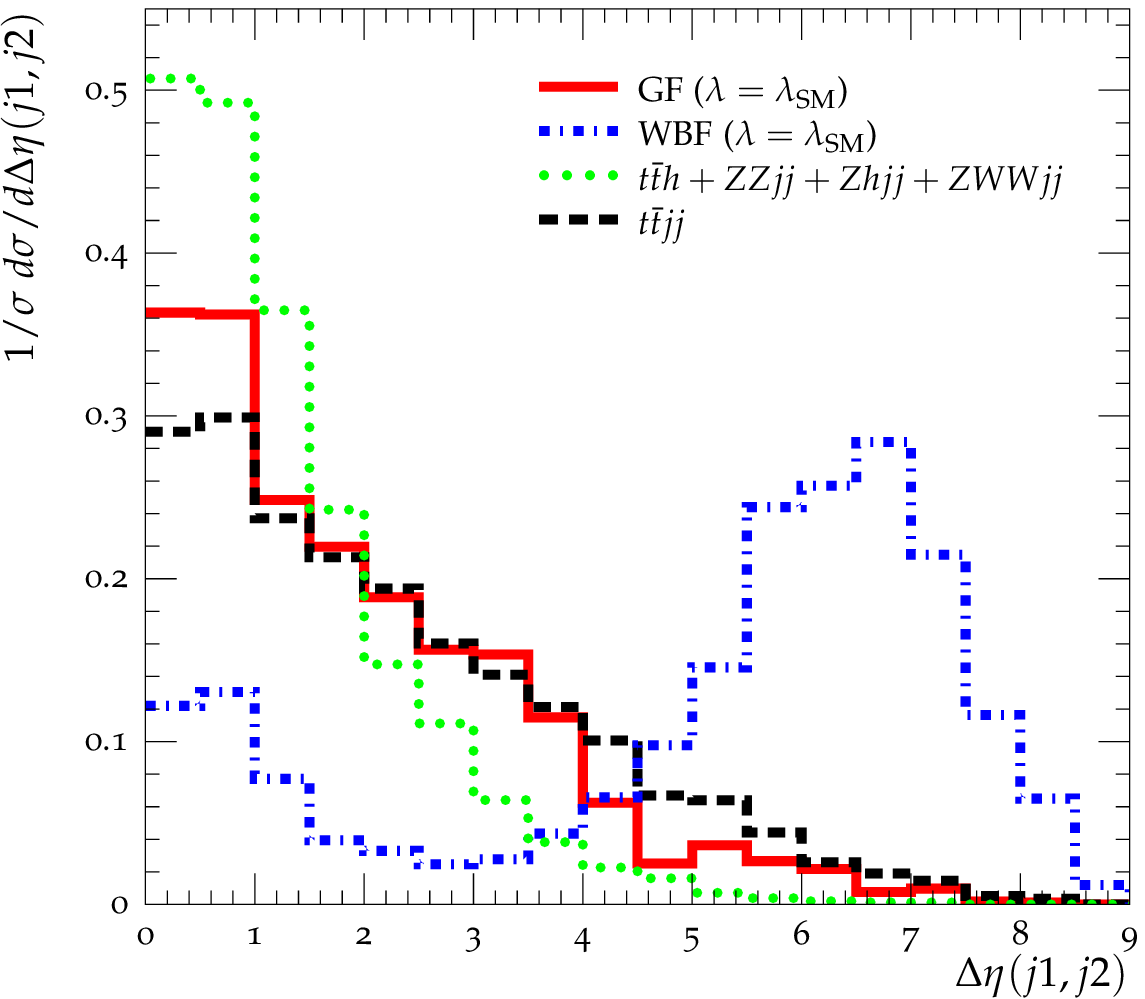}
\hfill
\includegraphics[width=0.45\textwidth]{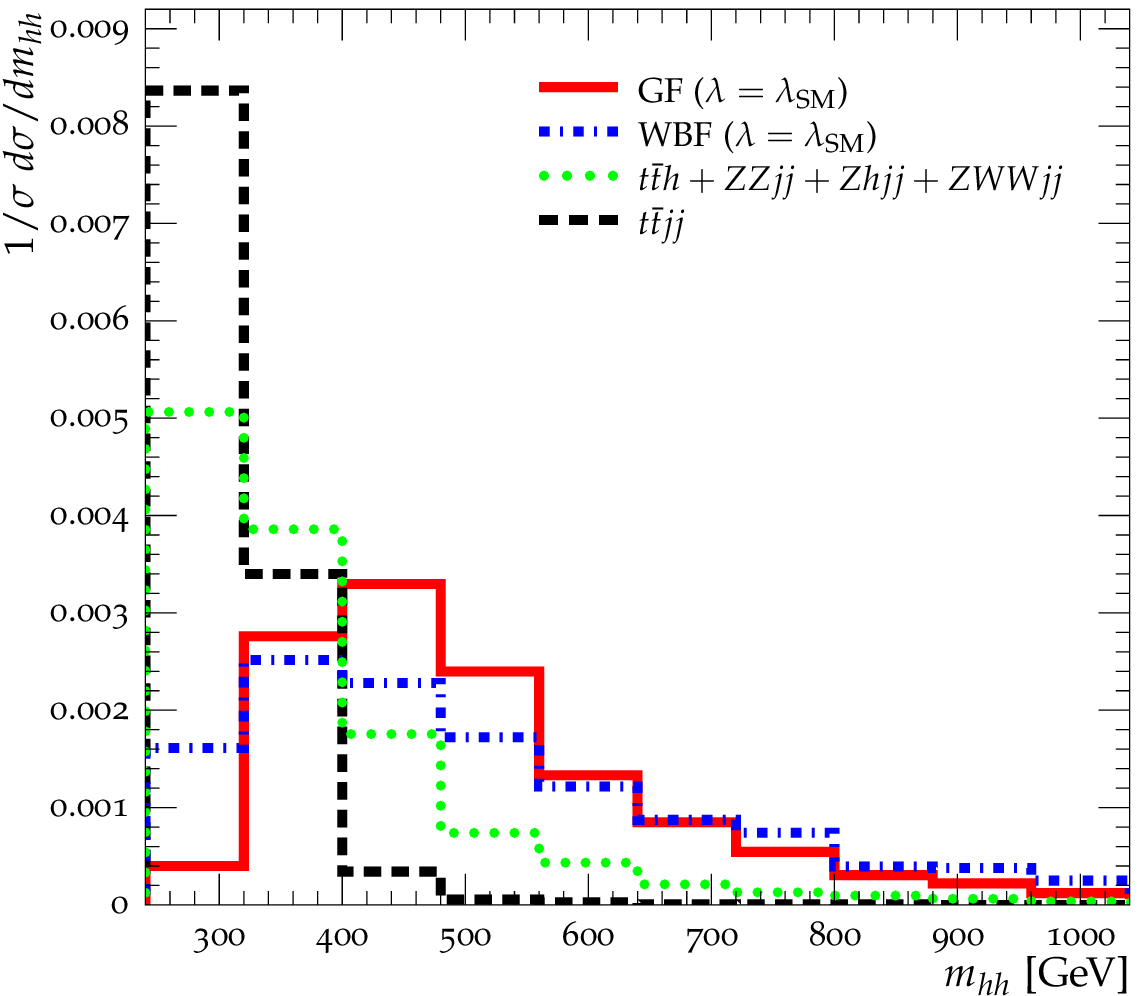}
\caption{Shape comparison of $\Delta \eta(j1,j2)$ and $m_{hh}$
  distributions for our two sources of signal (GF and WBF), the dominant background
  $t\bar{t}jj$ and the rest of the backgrounds (stacked scaled by
  relative cross sections), after the Base Selection of
  Section~\ref{sec:tt} has been applied.}
\label{fig:basehadronlvl}
\end{figure*}

We begin the hadron-level analysis implemented in
Rivet~\cite{Buckley:2010ar} by recreating a base selections similar
to~\cite{hhjj}:\footnote{Our analysis has been validated with two
  independent implementations.}
\begin{enumerate}[1.)]
\item We require two tau leptons using a two tau-trigger based on
  staggered transverse momentum selection cuts $p_T \ge
  29,20~\text{GeV}$ in $|\eta_\tau|<2.5$ and assume a flat tau tagging
  efficiency of 70\% with no fakes.

  Jets are constructed by clustering $R = 0.4$ anti-$k_T$ jets using
  \textsc{FastJet}~\cite{Cacciari:2011ma} with $p_{T,j} \ge 25$ GeV
  and $| \eta_j | \le 4.5$.
\item The two leading jets are $b$-tagged with an acceptance of 70\%
  and fake rate of 1\% \cite{btagging} in the central part of the
  detector $|\eta_j|<2.5$. We remove events if either of the two
  leading jets overlaps with a tau. Any additional jets which do not
  overlap with a tau are considered as potential ``tagging jets'', of
  which we require at least two.
\item As a final step of this base selection we require the $b$ jet
  and tau pairs to reproduce the Higgs mass of 125 GeV within $\pm 15$
  and $\pm 25~\text{GeV}$ respectively.\footnote{A high mass
    resolution is a crucial cornerstone of any successful di-Higgs
    analysis to assure a minimum pollution of $Z$ boson decay
    backgrounds~\cite{us2}.} 
\end{enumerate}
The signal and background cross sections after these cuts
are presented in the Base Selection column of
Table~\ref{tab:xstable}.  We find that the background contribution of
$t\bar{t}jj$ dominates with $t\bar{t}h$ also providing a larger-than-signal
background resulting in $S/B\sim~1/300$, making a study based only on these 
selections extremely challenging. Since we only have $\sim 40$ expected 
gluon fusion and $\sim~10$ expected weak boson fusion events at 3 ab$^{-1}$ 
luminosity, additional selections must also be careful to retain enough signal
cross section to allow statistically meaningful statements to be made
with a finite amount of data.

Shape comparisons for the rapidity and dihiggs invariant mass distributions  as
motivated in the previous section are shown in
Fig.~\ref{fig:basehadronlvl}. Indeed, as expected, cutting on the
angular distance of the jets will serve to both purify towards a
WBF-only selection at a reduced background rate. The dominant
backgrounds are unlikely to produce a large invariant mass
$m_{hh}$. However the WBF contribution, due to the lack of the $2m_t$
threshold peaks at a considerably lower invariant mass, leading to
significant decrease of the WBF contribution for a reasonably strong
cut on $m_{hh}$, which is required to observe the $hhjj$ signal at the
given low signal yield, even at 3 ab$^{-1}$ luminosity.

\begin{figure*}[t!]
  \centering
  \subfigure[\label{fig:nosystematics}]{\includegraphics[width=0.45\textwidth]{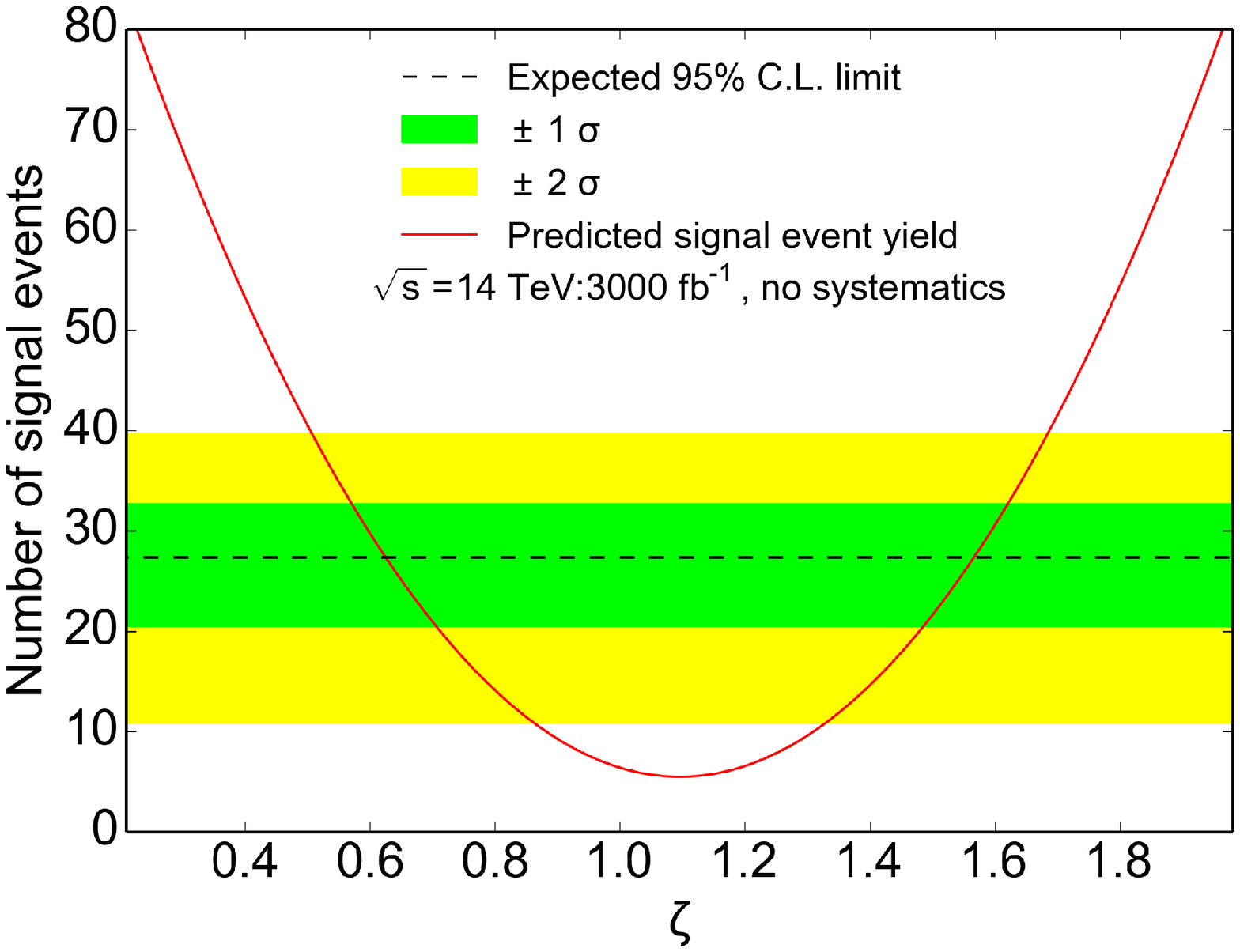}}
  \hfill
  \subfigure[\label{fig:20systematics}]{\includegraphics[width=0.45\textwidth]{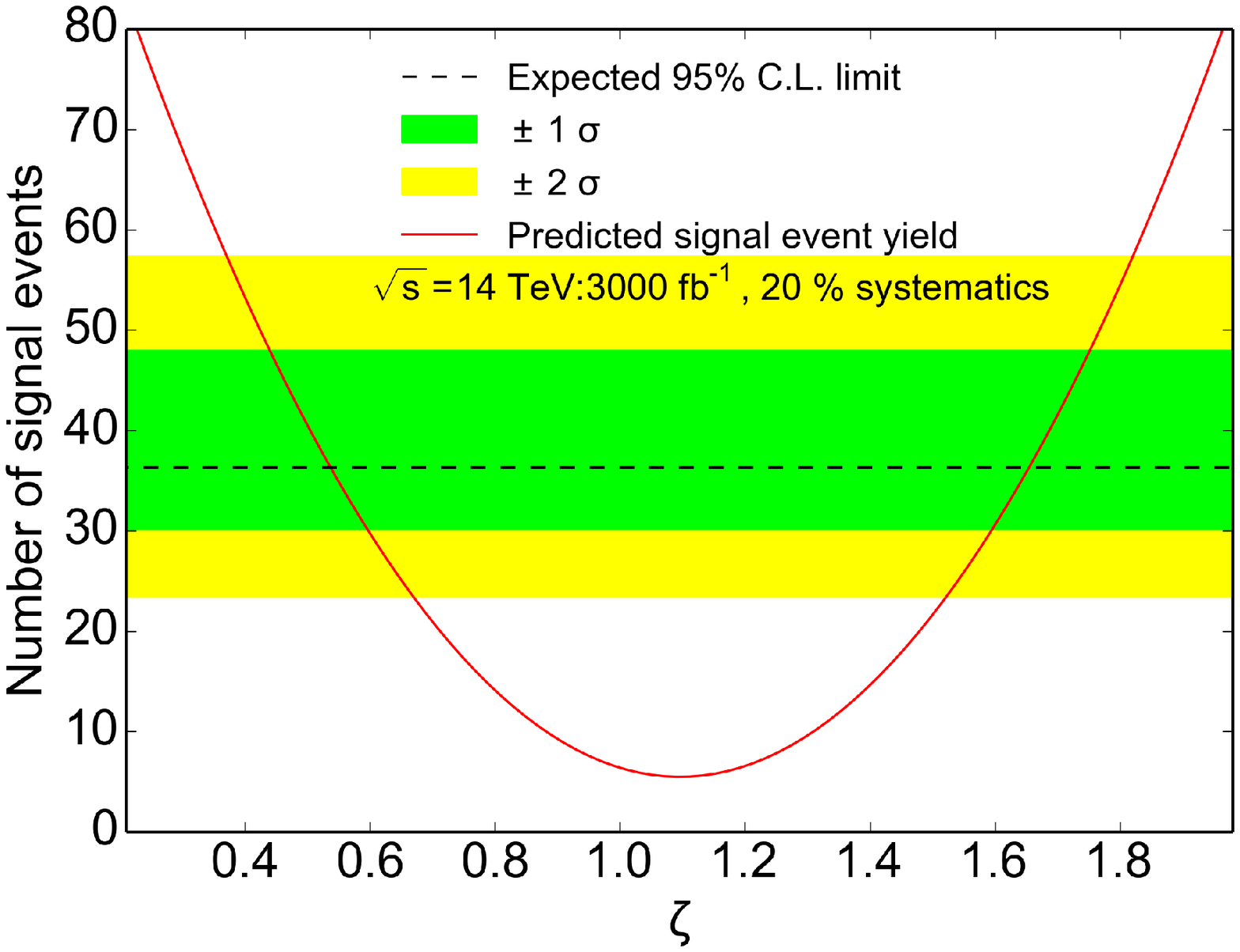}}
  \caption{Expected limits on the gauge-Higgs quartic couplings $\zeta=g_{VV hh}/g_{VV hh}^{\rm{SM}}$ under the assumption of no systematic 
    uncertainties~(a) and 20\% systematic uncertainties~(b).}
  \label{fig:zetalimits}
\end{figure*}

\begin{figure*}[t!]
  \centering
  \subfigure[\label{fig:gfjetti32}]{\includegraphics[width=0.45\textwidth]{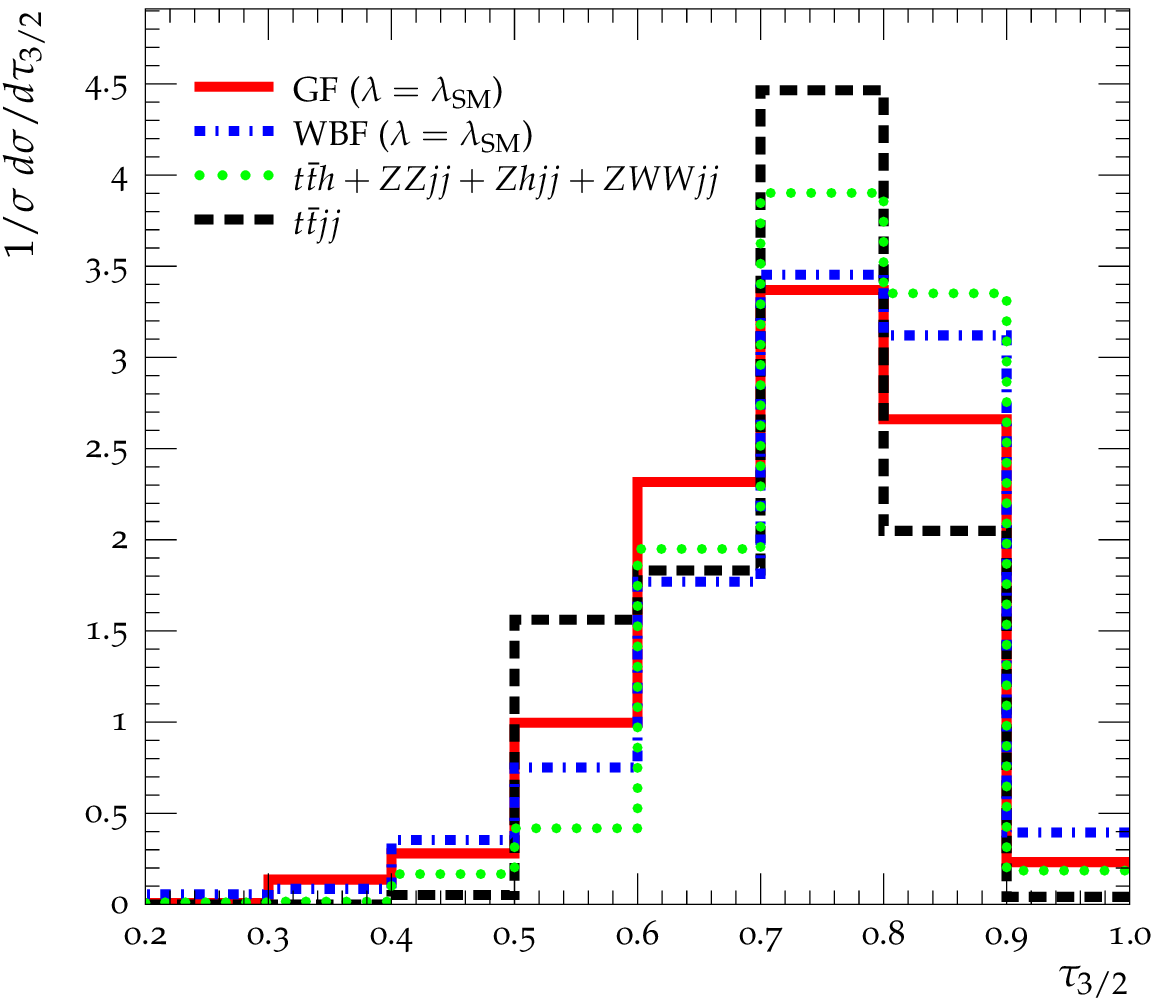}}
  \hfill
  \subfigure[\label{fig:wbfjetti32}]{\includegraphics[width=0.45\textwidth]{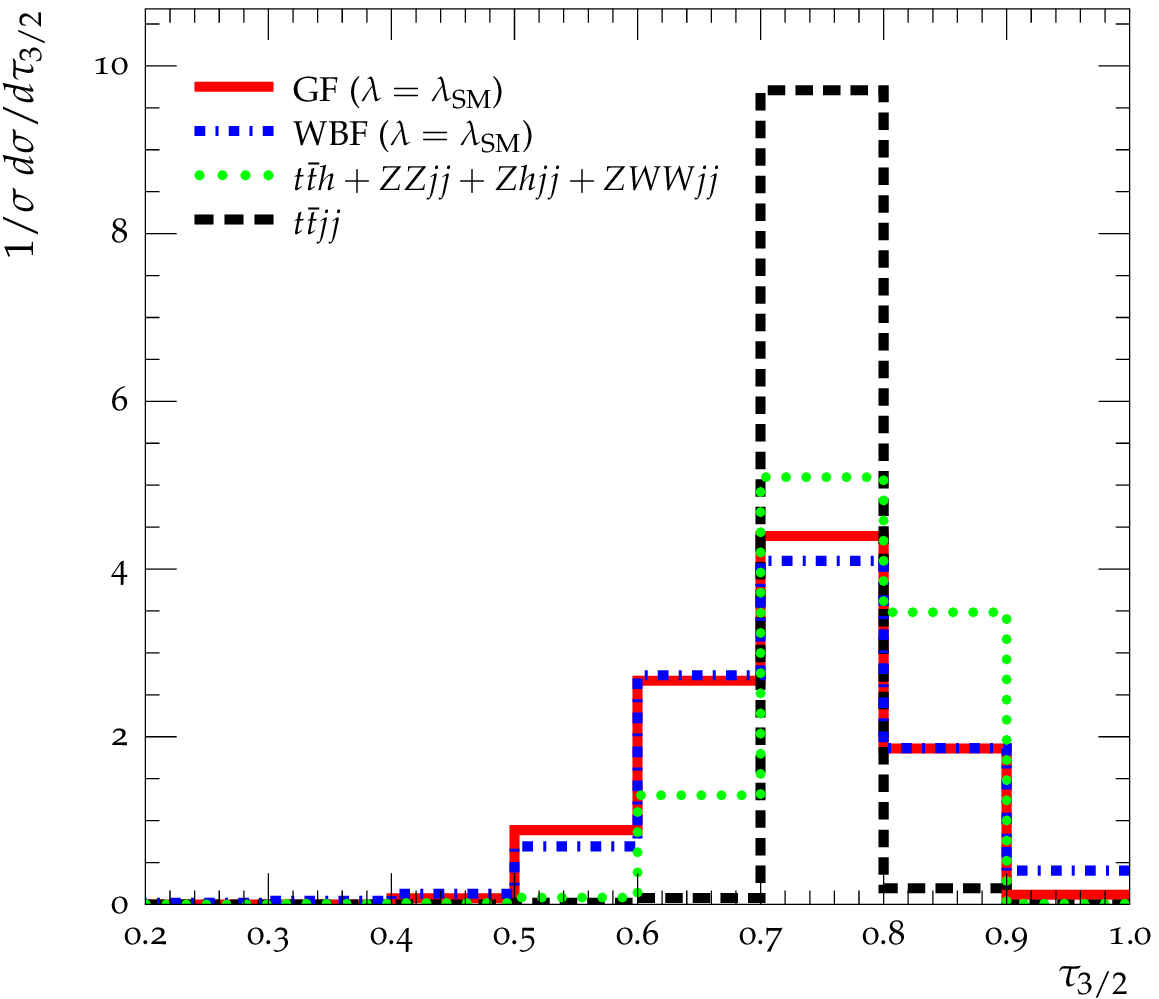}}\\[0.2cm]
  \subfigure[\label{fig:gfthrustmaj}]{\includegraphics[width=0.45\textwidth]{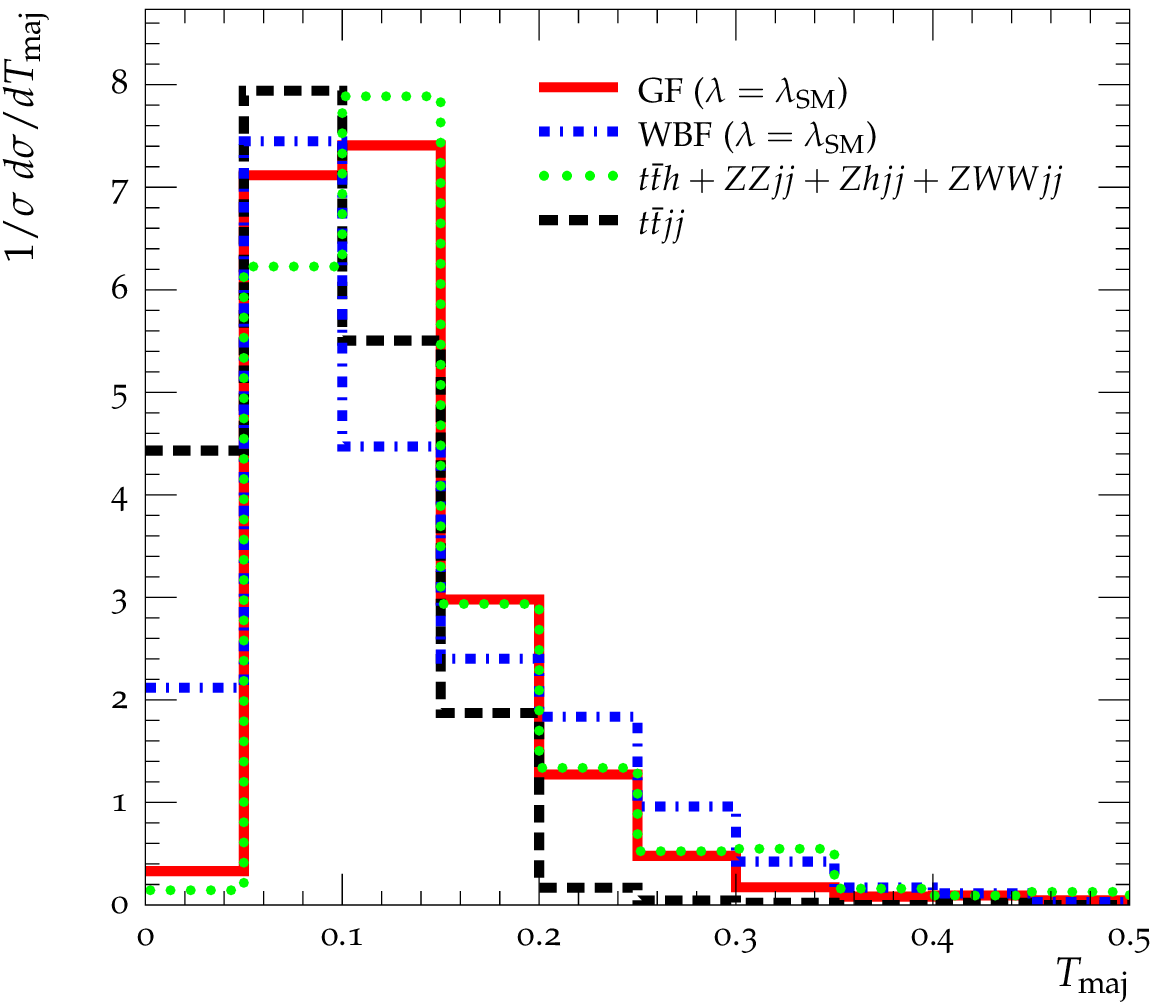}}
  \hfill
  \subfigure[\label{fig:wbfthrustmaj}]{\includegraphics[width=0.45\textwidth]{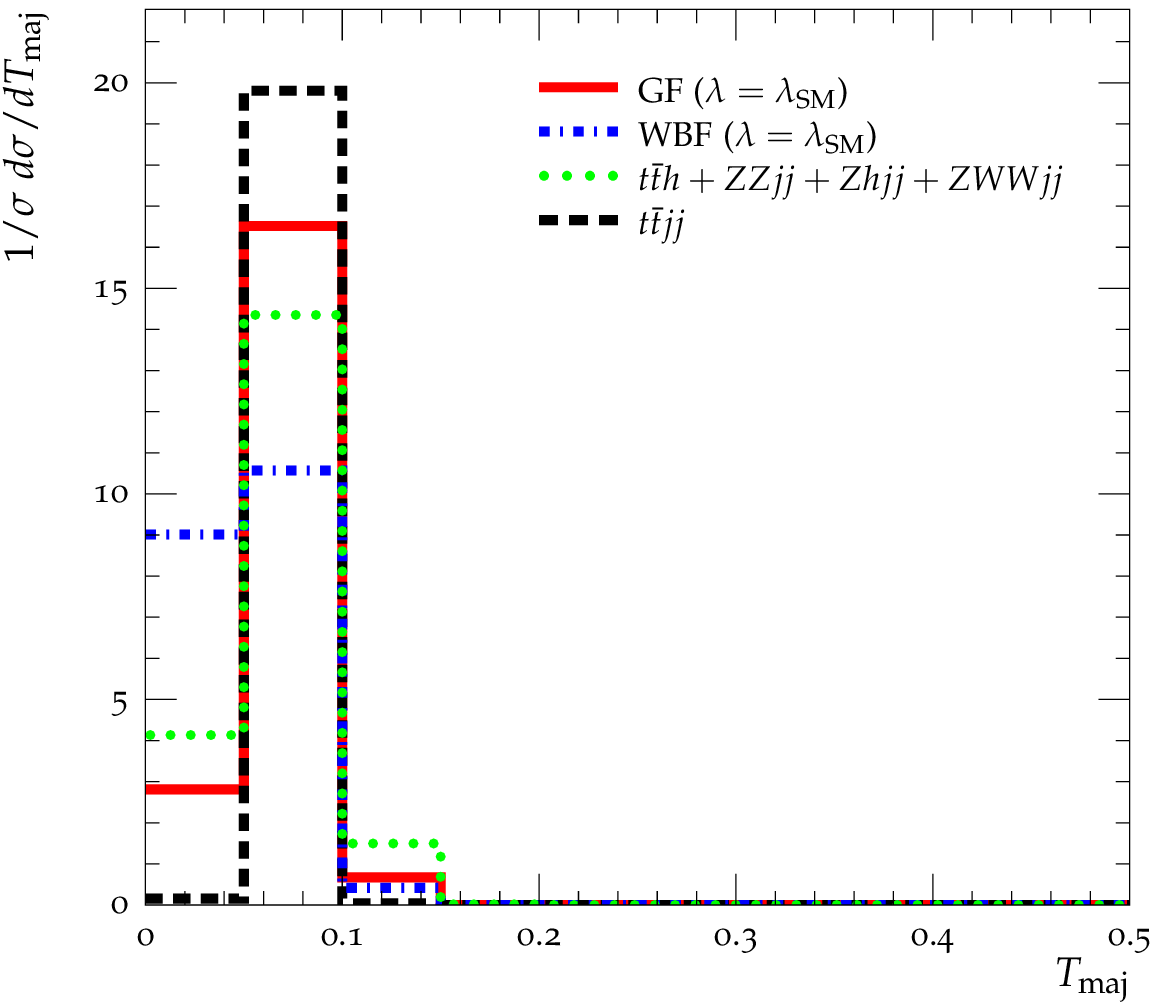}}
  \caption{Shape comparisons of $N$-jettiness and thrust calculated in the
    major direction after the gluon fusion selection of Sec.~\ref{sec:isolgf} (a,c)
    and WBF Selection of Sec.~\ref{sec:isolwbf} (b,d) have been applied.}
   \label{fig:evtshape}
\end{figure*}

\subsection{Prospects to isolate gluon fusion}
\label{sec:isolgf}
We can extend the analysis outlined in Sec.~\ref{sec:tt} with the aim
to purify the selection towards the GF component.\footnote{Following
  the analysis of \cite{Bredenstein:2008tm}, we can expect negligible
  interference between WBF and GF and which allows us to make this
  distinction.} We make use of the hard Higgs candidates to greatly
reduce the backgrounds by requiring $m_{hh} \ge 500$ GeV and
additionally require $\Delta \eta(j1,j2) \le 5$ to minimise the weak
boson fusion contribution.  The signal and background cross sections
after these cuts are applied are presented in the `GF Selection'
column of Table~\ref{tab:xstable}.

The total background is reduced by a factor of $\sim~100$ while the
gluon fusion contribution only is reduced by a factor of $\sim 2.5$
which allows for an encouraging $S/\sqrt{\text{B}}~\sim~1.7$ with
3 ab$^{-1}$ of data. The weak boson fusion contribution is 
also suppressed compared to GF which allows for a clean probe of the 
physics accessible in the gluon fusion contribution.

\subsection{Prospects to isolate weak boson fusion}
\label{sec:isolwbf}
Similarly we can extend the analysis towards isolating the WBF
component. Since it has slightly softer Higgs candidates we require
$m_{hh} \ge 400$ GeV and $\Delta \eta(j1,j2) \ge 5$ to reduce both the
gluon fusion and background contributions.  The signal and background
cross sections after these cuts are applied are presented in the `WBF
Selection' column of Table~\ref{tab:xstable}.

The total background is reduced by a factor of $\sim~300$ while three
times more of the weak boson fusion contribution is retained compared
to the GF selection, resulting in $S/\sqrt{{B}}~\sim~0.8$ with 3
ab$^{-1}$ of data due to the large reduction in the gluon fusion
contribution. However even so the WBF selection is composed of one-to-three parts
GF to WBF, which means measurements of physics that only enters the weak
boson fusion contribution will need to take this gluon fusion
``pollution'' into account.

\subsection{Constraining the quartic $VV hh$ contribution}
\label{sec:limitzeta}
As mentioned in Section~\ref{sec:wbf} there is a contribution from
quartic $VV hh$ vertices to the WBF induced signal, and modifications
of the corresponding $g_{VV hh}$ couplings away from their SM values
using the Higgs Cross Section Working Group $\kappa$ framework
\cite{Heinemeyer:2013tqa} will greatly enhance the signal cross section. This
allows us to constrain $\zeta$ defined by $g_{VV hh} = \zeta \times
g_{VV hh}^\text{SM}$.  To achieve this we have generated events with
varying $\zeta$ using \textsc{MadEvent} v5 and applied the WBF
selections described in Section~\ref{sec:isolwbf} to estimate the
enhancement of the signal, which is compared to expected cross section
limits on the signal with 3 ab$^{-1}$ of data in the WBF selection
under the assumptions of no systematic uncertainties and 20\% total
systematic uncertainties for comparison. The results are presented in
Figure~\ref{fig:zetalimits}.  We find that in the more realistic
scenario of 20\% systematic uncertainties the expected constraint on
the $g_{VV hh}$ couplings is $0.55 < \zeta < 1.65$ at 95\% confidence
level. A measurement of $pp\to hhjj$ is therefore crucial to constrain
new physics which enters predominantly through enhancements to $g_{VV
  hh}$.

\subsection{Event shapes of the tagging jets system}
The analysis strategies outlined so far have mainly relied on
exploiting correlations in the di-Higgs system, with only $\Delta
\eta(j1,j2)$ carrying information about the tagging jets. Following similar applications in the context of single Higgs
production \cite{Englert:2012ct}, we investigate a range of event
shapes in the tagging jets system in the following, which could offer
additional discriminating power through capturing colour correlations
in the different signal contributions beyond angular
dependencies. More specifically, we will focus on $N$-jettiness
\cite{jettiness,Thaler:2010tr} and thrust major which provided 
the best results.

We calculate $N$-jettiness by minimising
\begin{equation}
 \tau_N = C \sum_k p_{T,k} \min(\Delta R_{k,1},\dots, \Delta R_{k,N})
\end{equation}
where $C$ is a normalisation which cancels when taking the ratio of
two $\tau$s, the sum is taken over all visible momenta which do not
belong to one of the identified Higgs candidates within $|\eta| < 5$,
and $\Delta R_{k,n}$ is the distance in the $\eta - \phi$ plane 
between the $k$-th momentum and the $n$-th reference vector. 
$\tau_{3/2}$ is then explicitly given by $\tau_{3}/\tau_{2}$.

Thrust major is defined by
\begin{subequations}
  \begin{equation}
    T_{\text{maj}}=\max_{{\vec{n}}\cdot{\vec{n}}_T=0} \frac{\sum_k
      |{\vec{p}}_k\cdot {\vec{n}}|}{\sum_k |{\vec{p}}_k|}
  \end{equation}
  where ${\vec{n}}_T$ is the normalised thrust vector
  \begin{equation}
    {\vec{n}}_T=\max_{\vec{n}}\frac{\sum_k
      |{\vec{p}}_k\cdot {\vec{n}}|}{\sum_k |{\vec{p}}_k|}\,,
  \end{equation}
\end{subequations}
Again the sums run over all visible momenta which do not belong
to one of the identified Higgs candidates within $|\eta| < 5$.

We find $\tau_{3/2}$ and $T_\text{maj}$ show promise for improving the
WBF selection, but the signal cross section is already too low for us
to be able to make meaningful use of this insight. The $\tau_{3/2}$
and $T_\text{maj}$ distributions after the GF and WBF selections have
been applied are presented in Fig.~\ref{fig:evtshape}. Cutting, e.g.,
on $T_{\text{maj}}<0.05$, the gluon fusion contribution is reduced by
80\%, while the WBF contribution is reduced by only 55\% amounting to
a total of 2 expected WBF and 0.3 expected GF events, with backgrounds 
very strongly suppressed. This means 
that WBF can in principle be observed at a small rate that can be used 
to set constraints on new physics in an almost GF-free selection with 
greatly reduced backgrounds.

The event shape distributions can also be used to greatly reduce the
background in the GF selection, Fig.~\ref{fig:gfthrustmaj}. It should
be noted that these improvements of GF vs WBF vs background ultimately
depend on underlying event and pile up conditions and have to be taken
with a grain of salt at this stage early in run 2. However the clear
separation that can be achieved with these observables indicate that
an analysis employing MVA techniques could, at least in theory,
significantly improve the results presented here. These techniques may
also prove useful at a 100~TeV collider where the dihiggs production
cross-section is substantially higher~\cite{Barr:2014sga}.

\section{Summary and Conclusions}
After discovering single Higgs production at the Large
Hadron Collider, new analysis strategies need to be explored to
further constrain the presence of new physics beyond the Standard
Model. Higgs pair production is pivotal in this regard as constraints
from multi-Higgs production contain complementary information, in
particular with respect to the Higgs boson's self-interaction. Cross
sections for di-Higgs production are generically small at the LHC,
which highlights the necessity to explore other viable channels than
$pp\to hh$ to enhance sensitivity in a combined fit at high
luminosity. To this end, we have investigated $pp\to hh jj$ production
in detail in this paper. Keeping the full top and bottom mass
dependencies, we find sensitivity of $pp\to hhjj$ searches at the LHC
for production in the SM and beyond. The gluon fusion contribution
remains important at high invariant di-Higgs masses where the dominant
backgrounds can be suppressed to facilitate a reasonable signal vs
background discrimination. Unfortunately, the gluon fusion
contribution remains large even for selections that enhance the weak
boson fusion fraction of $pp\to hhjj$ events. This ``pollution'' is
important when such selections are employed to set constraints on new
physics effects that enter in the WBF contribution exclusively. Large
new physics effects in the WBF contribution can still be constrained,
which we have illustrated through an investigation of the constraints
that can be set on deviations of the quartic $VV hh$ couplings from
their SM values with the HL-LHC, demonstrating that a measurement of
$pp\to hhjj$ will provide a powerful probe of these.  Employing
observables which are intrinsically sensitive to the different colour
correlation of WBF compared to GF, the discrimination between GF, WBF,
and background can be further improved.  However, the signal cross
section is typically already too small to use such a strategy to
constrain the presence of new physics if those effects are only a
small deviation around the SM. If new physics effects are sizable,
such an approach will remain a well-adapted strategy to minimise GF
towards a pure WBF selection.

\vskip 1\baselineskip
\noindent {\emph{Acknowledgements.}} CE and MS are grateful to the
Mainz Institute for Theoretical Physics (MITP) for its hospitality and
its partial support during the workshop ``Higgs Pair Production at
Colliders'' where part of this work was completed. CE is supported in
parts by the Institute for Particle Physics Phenomenology
Associateship programme. KN thanks the University of Glasgow College
of Science \& Engineering for a PhD scholarship. This research was
supported in part by the European Commission through the ’HiggsTools’
Initial Training Network PITN-GA-2012-316704.


\end{document}